\documentclass[manuscript,screen]{acmart}

\usepackage{algorithmic}
\usepackage{algorithm}
\usepackage{subfigure}
\usepackage{graphicx}
\usepackage{url}
\usepackage{textcomp}
\usepackage{xcolor}
\newcommand{\RNum}[1]{\uppercase\expandafter{\romannumeral #1\relax}}
\usepackage{fancyhdr}
\usepackage{footnote}
\usepackage{natbib}
\usepackage{amsmath}

\AtBeginDocument{%
  \providecommand\BibTeX{{%
    \normalfont B\kern-0.5em{\scshape i\kern-0.25em b}\kern-0.8em\TeX}}}

\setcopyright{acmcopyright}
\copyrightyear{2018}
\acmYear{2018}
\acmDOI{XXXXXXX.XXXXXXX}

\begin{document}

\title{Accelerating Convolutional Neural Network by Exploiting Sparsity on GPUs}
\titlenote{New Paper, Not an Extension of a Conference Paper}

\author{Weizhi Xu}
\affiliation{%
  \institution{School of Information Science and Engineering, Shandong Normal University}
  \city{Jinan}
  \country{China}
}
\affiliation{%
  \institution{Dept. of Electrical \& Computer Engineering, University of Houston}
  \city{Houston}
  \country{USA}
}

\author{Yintai Sun}
\affiliation{%
  \institution{School of Information Science and Engineering, Shandong Normal University}
  \city{Jinan}
  \country{China}
}
\author{Shengyu Fan}
\affiliation{%
  \institution{School of Information Science and Engineering, Shandong Normal University}
  \city{Jinan}
  \country{China}
}

\author{Hui Yu}
\affiliation{%
  \institution{School of Information Science and Engineering, Shandong Normal University}
  \city{Jinan}
  \country{China}
}
\affiliation{%
  \institution{Dept. of Electrical \& Computer Engineering, University of Houston}
  \city{Houston}
  \country{USA}
}

\author{Xin Fu}
\affiliation{%
  \institution{Dept. of Electrical \& Computer Engineering, University of Houston}
  \city{Houston}
  \country{USA}
}
\authornote{Corresponding author}

\begin{abstract}
Convolution Neural Network (CNN) is an important deep learning method, which is widely used in many fields. However, it is very time consuming to implement CNN where convolution usually takes most of the time. There are many zero values in feature maps and filters, which leads to redundant calculations and memory accesses if dense methods are used to compute convolution. Many works recently make use of sparsity to skip the calculations for zero values to reduce the inference time of CNN. On the GPU platform, current works cannot fully exploit the sparsity of the feature map and achieve satisfactory performance. Therefore, we design a new parallel strategy to transform the feature map into a new storage format to avoid the redundant computation of zero-values on GPUs. Also considering the sparsity in the feature map, we propose a fused storage format to combine the convolution operation with the following pooling operation, in order to further improve the performance. 
We carry out experiments with mainstream CNN models and achieve better performance compared with cuDNN and cuSPARSE. For VGG-19, ResNet-50, DenseNet-121 and RegNetX-16GF, $1.97\times$, $2.23\times$, $2.74\times$ and $1.58\times$ speedups are obtained respectively over cuDNN. The speedups over cuSPARSE are $2.10\times$, $1.83\times$, $2.35\times$ and $1.35\times$ respectively when only using the first method.

\end{abstract}





\maketitle
\vspace{-5pt}
\section{Introduction} \label{introduction}
\vspace{-3pt}
Convolutional neural network (CNN) plays a crucial role in deep learning, which facilitates the development of computer vision \cite{EfficientNet}, medical image classification \cite{b1}, natural language processing \cite{b4}, recommender system \cite{b5} etc. CNN is also an essential part of many complex neural network models \cite{b6,b7}. However, it is time-consuming for training and inference of CNN. Improving the execution speed of CNN is of great importance for accelerating its applications. Convolution operations usually take up a large part of the total execution time of CNN, even more than 70\% in the well-known CNNs such as VGG, ResNet and YOLO \cite{b8,b39}.
Designing specific hardware architectures for convolution computation is a feasible way to accelerate CNN \cite{b9,b10}, e.g. Tensor Cores \cite{b11, STC}. Some algorithmic optimization techniques are also developed, such as Im2col \cite{b12,gemm_coordinated}, FFT \cite{b14}, and Winograd \cite{b15}. The above acceleration methods are integrated into cuDNN, which is a state-of-the-art library for deep learning on GPU \cite{b16}. 

The convolution operation in CNN refers to the process in which the convolution kernel samples on the feature map. 
The process of convolution contains a lot of multiplications and additions, so reducing these operations are effective for acceleration.
Network pruning \cite{CNN-Channel-Prune} and RELU activation are common operations in CNNs, which result in a large number of zero values in the network. For the feature map, the ratio of zero values evolves \cite{CompileSparse} and can be more than 0.8 in the deep layers of the network after multiple epochs. The pursuit of precision in CNNs leads to dozens of epochs, so the large sparsity in the feature map is inevitable. The calculation of these zero values, however, is useless for the convolution result \cite{cpu-sparseconv}. 
In other words, if we can skip the zero-value calculation in the convolution operation, this will reduce operations of multiplication and addition. 

Some efforts focus on reducing the calculations of zero values in neural networks by designing new hardware architecture \cite{b19, b20, b21,GoSPA,SNAP}, 
which obtain outstanding acceleration effects. However, new architectures based on FPGA or ASIC have a relatively long development cycle, compared with CPU or GPU.
Some algorithms exploiting sparsity for convolution operations on CPU are proposed and achieve considerable speedup over their dense counterparts  \cite{cpu-sparseconv,b18,cpu-cnn-Park,b22,amer2021high}. GPU is widely used for accelerating CNN because of its strong computing power for parallelizable processes such as large-scale matrix multiplication based on highly-optimized cuBLAS \cite{im2col2,sc2016}.
Current implementations of sparse convolution on GPU are usually based on sparse libraries such as cuSPARSE \cite{GPU-arXiv-2020, GPU-ICPP-2019}. For the convolution layer, feature maps and filters can be processed with three steps, extension, compression and sparse matrix computation by cuSPARSE.

However, exploiting sparsity in convolution operation can hardly achieve satisfactory performance when CNN is implemented on GPU \cite{GPU-NIPS-2015}, and very limited speedup over cuBLAS-based convolution methods is obtained \cite{banlanced-sparsity, STC}. The reasons are as follows.
1) The above three steps are separately completed on GPU, so the data are loaded from and stored into global memory at least three times. When using cuSPARSE, the data are even needed to be transferred between CPU and GPU. 
2) The cost of real-time compression affects the performance.
3) Sparse matrix computation is less efficient than its dense counterparts on GPUs  \cite{GPU-Escort-2018}.

On the other hand, the pooling operation usually starts after the convolution results are transferred from GPU to CPU in traditional CNN implementations. This increases traffic not only between global memory and shared memory on GPU but also between CPU and GPU. To address this problem,
some works \cite{Pytorch,b37,Fused-layer, Data-Reuse,MegaKernels} consider integrating different CNN layers into one GPU
kernel to decrease communication overhead. However, the adjacent layers are simply put together in one kernel, and the sparsity for convolution operation is not considered.
In this paper, we propose two novel methods to accelerate CNN on GPUs. Our contributions can be summarized as follows.

\begin{itemize}
    \item We propose a novel storage format (ECR, Extended and Compressed Row) for sparse feature maps. We design a convolution algorithm based on ECR, which calculates convolution by skipping zero values and reduces the amount of computation. Furthermore, the ECR method can complete extension, compression and sparse matrix computation by only accessing global memory once.
    \item We also propose another storage format (PECR, Pooling-pack Extended and Compressed Row). Besides exploiting sparsity, convolution layer and pooling layer are computed together, which effectively reduces not only the traffic between CPU and GPU but also the traffic from off-chip memory to on-chip memory of GPU.
    \item We evaluate the proposed algorithms on the GPU platform. The result shows that our methods can achieve the state-of-the-art speedup over cuDNN. For VGG-19, ResNet-50, DenseNet-121 and RegNetX-16GF, $1.97\times$, $2.23\times$, $2.74\times$ and $1.58\times$ speedups are obtained respectively over cuDNN. The speedups over cuSPARSE are $2.10\times$, $1.83\times$, $2.35\times$ and $1.35\times$ respectively when only using ECR.
\end{itemize}

The rest of this paper is organized as follows. Section  \ref{background} presents the background. Section \ref{Related work} discusses the related work. Section \ref{motivation} introduces the motivation. Section \ref{ecr} describes the proposed ECR format and its convolution method. Section \ref{pecr} describes the PECR format and the corresponding convolution and pooling method. Section \ref{discussion} introduces Optional Convolution and Pooling Algorithm (OCPA) which can use both ECR and PECR for an entire network. Section \ref{experiments} evaluates the proposed methods on several CNN models. Section \ref{conclusion} concludes this paper.

\section{Background} \label{background}
In this section, we first introduce  Graphic Processing Units (GPUs) and Compute Unified Device Architecture (CUDA). Then, we present the convolution operation and the pooling operation for CNN. The notations used in the paper are described in Table \ref{notation}.

\begin{table}[h]
\vspace{-10pt}
	\centering
	\caption{Notations in the paper}\label{notation}
	\begin{tabular}{|c|c|c|c|}
		\hline
		$B_i$ & block id of GPU thread block & $T_i$ & thread id of GPU thread \\
		$i_w$ & width of feature map & $i_h $ & height of feature map \\
		$k_w$ & width of the kernel & $ k_h$ & height of the kernel\\
		$p_w$ & width of pooling window & $p_h $ & height of pooling window \\
		$c_s $ & stride size for convolution & $ p_s$ & stride size for pooling \\
		\hline
	\end{tabular}
\vspace{-20pt}
\end{table}

\begin{figure}[h]
	\centering
	\includegraphics[scale=0.4]{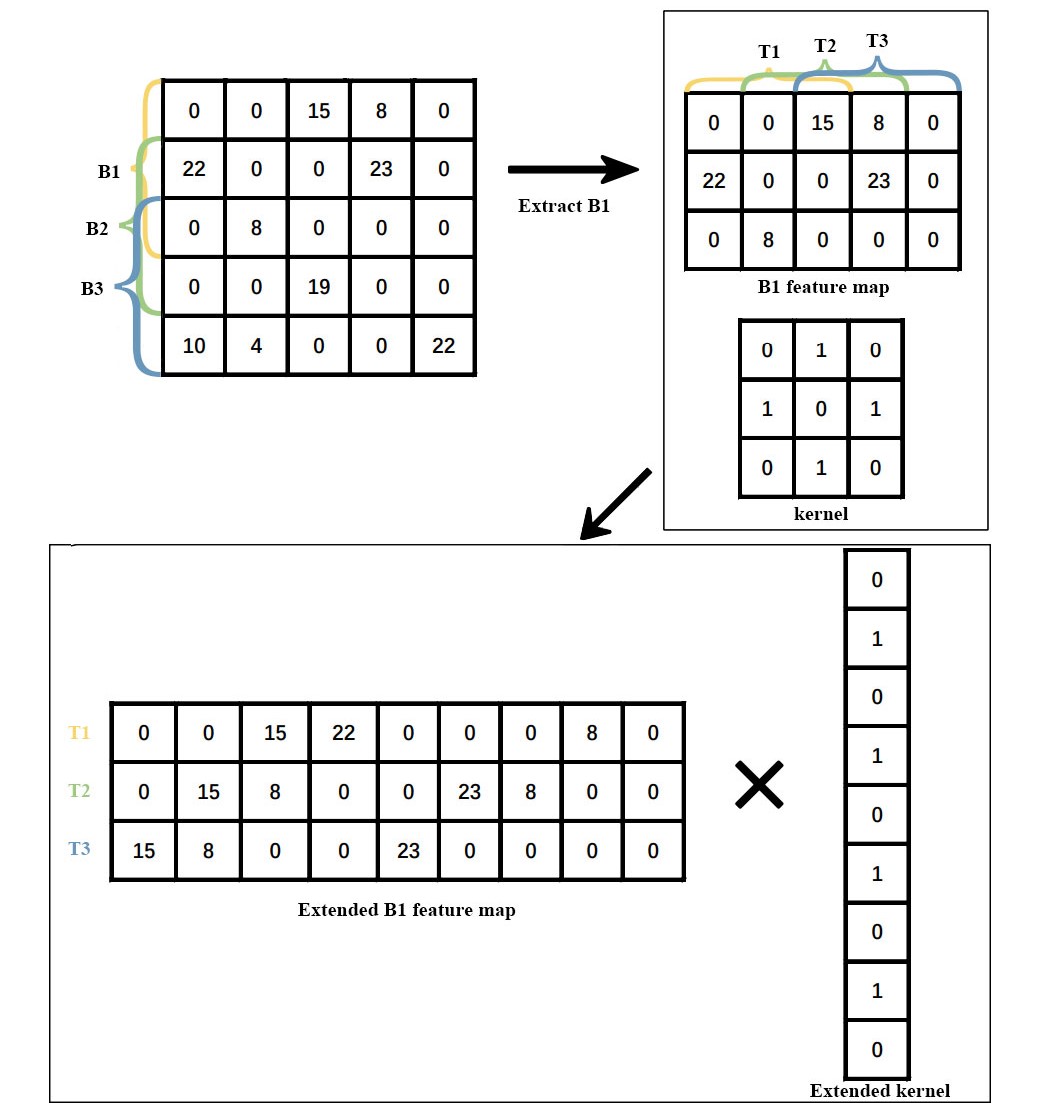}
	\caption{Example of convolution computation by matrix-vector multiplication.}
        \label{im2col}
\end{figure}

\subsection{GPU/CUDA platform}
GPUs are widely used to meet the needs of high performance computing for many applications, such as scientific computing, deep learning, Bioinformatics \cite{biogpu}, and so on.
One GPU usually contains multiple Streaming Multiprocessors (SMs), and each SM contains multiple Streaming Processes (SPs) \cite{nvidiatesla}. Each SM can access a register file that works at the same speed as SP. Usually, each SM contains an L1 cache, and all SMs shared an L2 cache. Each SM also has a shared memory that is similar to the L1 cache, but its data replacement is controlled by the programmer.
The shared memory is shared between SPs within the same SM. 
Global memory is off-chip and can be used for data transfer between GPU and CPU. CPU can access the global memory through the PCI-E or AXI bus. The latency of transferring data from CPU to GPU and from global memory to shared memory is very high, so it is beneficial to put reusable data in shared memory instead of frequently visiting CPU memory or global memory.
CUDA is a programming platform designed for GPU architecture. CUDA makes parallel programming on GPU more acceptable and promotes the development of parallel applications. On CUDA platform, all threads are contained in a thread grid, which consists of multiple thread blocks. The threads in a thread block share the same shared memory space. According to GPU's characteristics, threads are organized in the form of warp, which generally includes 32 threads \cite{era, nvidiacuda}.

\subsection{Convolution on GPU} 
Convolution calculations have been converted to GEneral Matrix Multiplications (GEMM) on GPU \cite{b12}. 
Using GPU platform to calculate GEMM has been deeply studied \cite{MM_gpu, gemm_coordinated}.
Im2col is a common way to compute convolutions in CNN frameworks such as Caffe \cite{Caffe}, TensorFlow \cite{Tensorflow}, PyTorch \cite{Pytorch} and cuDNN \cite{b12}.
This method needs to transform the input feature maps $I$ into an extended matrix $B$, so that the convolution results can be obtained from the GEMM:
$$O = CONV(F, I) = A \cdot B = A \cdot Im2col(I),$$
where $O$ is output of convolution; $F$ is convolution filters; $A$ is a matrix contains the filters $F$; $B$ is the result of applying the Im2col transform to the input feature map $I$ according to the filter size and stride.

As shown in Fig. \ref{im2col}, the feature map is divided into three large convolution block rows ($B1$, $B2$ and $B3$), each of which corresponds to one row of final convolution results. One thread block of GPU is assigned to compute one convolution block row. In each convolution block row, convolution kernel moves horizontally. Therefore, each convolution block row is divided into three convolution windows, each of which corresponds to one convolution result. Each convolution result is computed by one GPU thread  ($T1$, $T2$ or $T3$). The values in the convolution window are extended to one row of a matrix, and the convolution kernel is extended to a vector. In this way, the convolution calculation is converted to matrix-vector multiplication. The result of the matrix-vector multiplication corresponds to a row of convolution results. Matrix-vector multiplication is much faster than convolution operation on GPUs, so the calculation speed is substantially improved. Accordingly, convolution operation with multiple convolution kernels can be transformed to matrix-matrix multiplication, which is also very suitable for GPU computing.
In recent years, there have been some optimization methods for convolution on GPUs, which will be discussed in related work.

\subsection{Pooling operation}
To reduce dimensions of the output from convolution layer and prevent overfitting, the pooling layer is added after the convolution layer in most CNNs.  CNNs usually have three kinds of pooling operations, mean-pooling, max-pooling, and stochastic-pooling.
Similar to the convolution operation, the pooling operation also has a sliding window on the feature map, but no multiplication operation is required in this window. In the sliding window, the mean-pooling takes the average value from all values in the window. The max-pooling takes the max value from all values in the window. The stochastic-pooling gives a probability to each value in the window, then the one with the highest probability is selected.

\begin{figure}[h]
	\centering
	
	\subfigure[VGG-11]{
		\begin{minipage}[t]{0.5\linewidth}
			\centering
			\includegraphics[width=1\textwidth,height=0.5\textwidth]{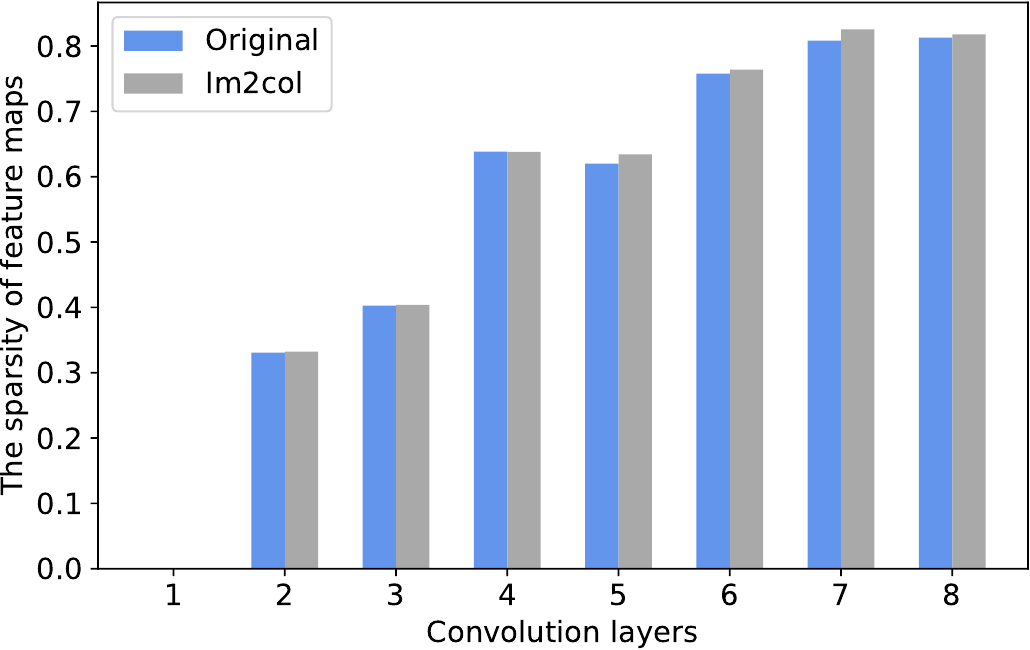}\\
			\vspace{0.02cm}
		\end{minipage}%
	}%
	\subfigure[VGG-13]{
		\begin{minipage}[t]{0.5\linewidth}
			\centering
			\includegraphics[width=1\textwidth,height=0.5\textwidth]{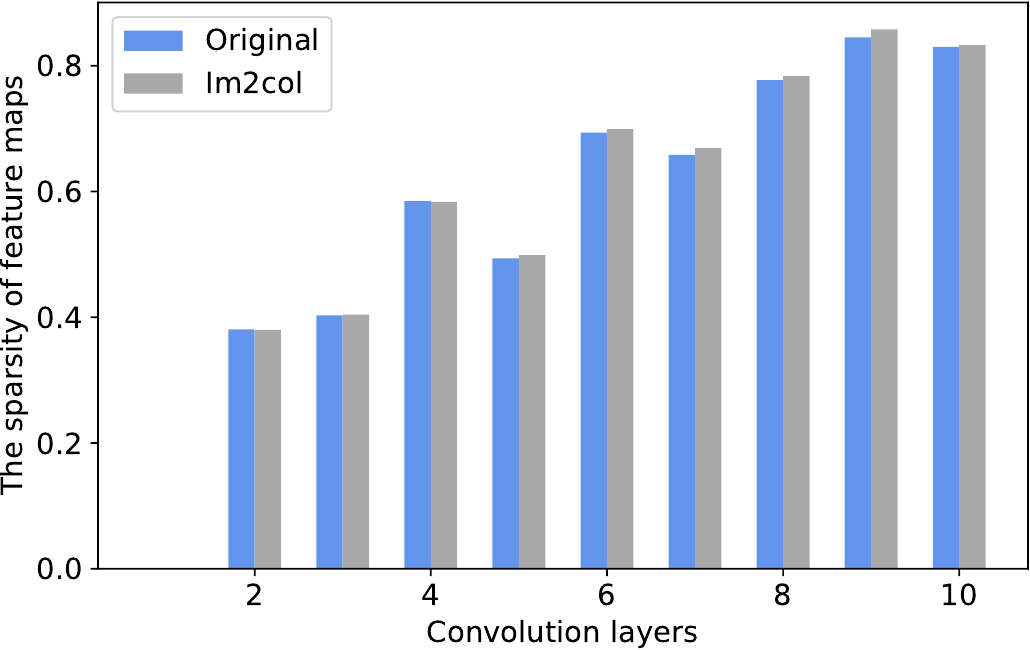}\\
			\vspace{0.02cm}
		\end{minipage}%
	}%

	\subfigure[VGG-16]{
		\begin{minipage}[t]{0.5\linewidth}
			\centering
			\includegraphics[width=1\textwidth,height=0.5\textwidth]{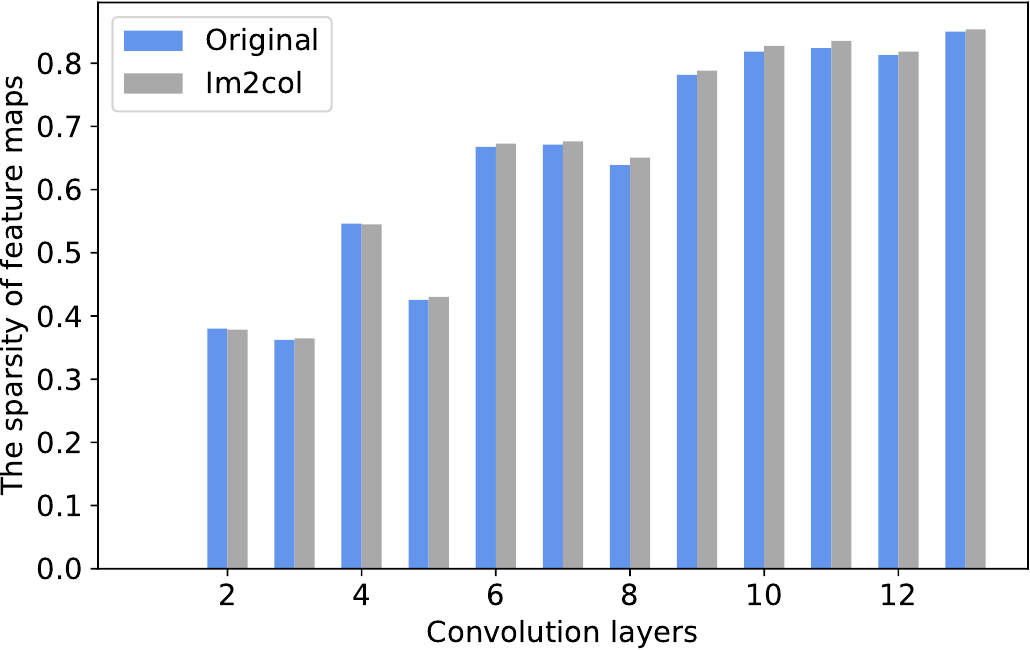}\\
			\vspace{0.02cm}
		\end{minipage}%
	}%
	\subfigure[VGG-19]{
		\begin{minipage}[t]{0.5\linewidth}
			\centering
			\includegraphics[width=1\textwidth,height=0.5\textwidth]{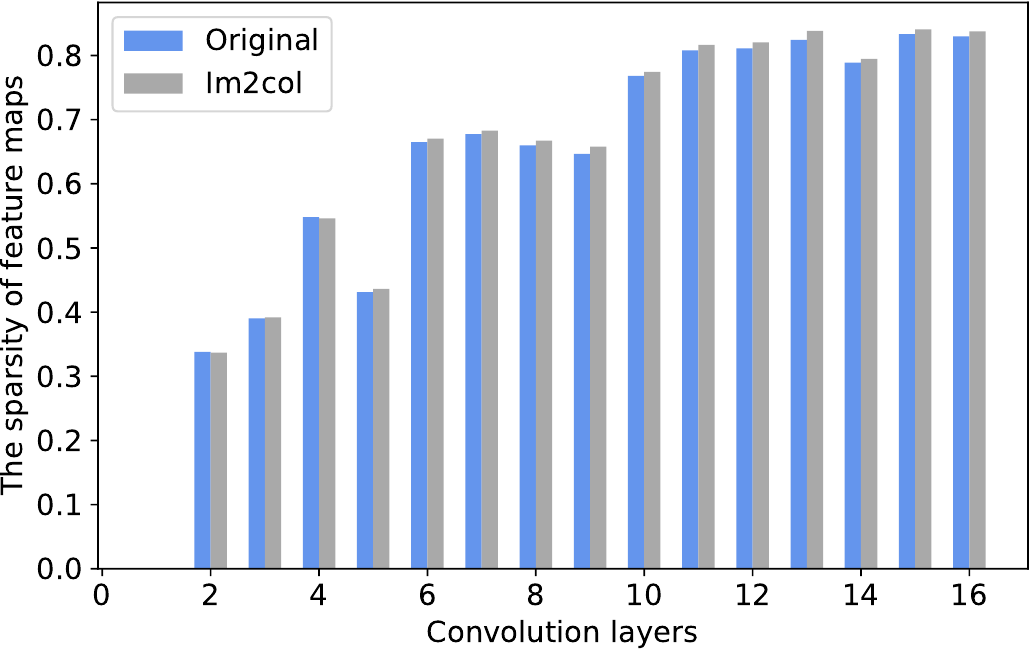}\\
			\vspace{0.02cm}
		\end{minipage}%
	}%
	\centering
	 \vspace{-10pt}
	\caption{The sparsity of the feature maps for VGG networks. The blue bar is the original feature map, and the grey bar is the feature map after matrix transformation as in Im2col. }
	
	\label{FigSparisity}
\vspace{-20pt}
\end{figure}

\section{Related work} \label{Related work}

CNN is an important kind of deep neural network, which can be applied in a broad range of domains. In this section, we summarize and classify the methods for accelerating CNNs.

\subsection{Dense methods for accelerating CNNs}

Dense methods for accelerating CNNs mainly refer to accelerating matrix multiplication for convolution with dense storage format. The dense methods can be divided into two categories, CPU or GPU, FPGA or ASIC.

%
\subsubsection{CPU or GPU}
Direct convolution needs less memory space than GEMM-based method but it is not efficient. Im2col first extends the feature maps and filters, and then transforms convolution into matrix multiplication. This method gets a good acceleration effect because GPU is good at accelerating GEMM \cite{3BLAS}. 
However, Im2col cannot achieve satisfactory performance for small feature maps because GEMM for small matrices cannot fully exploit the computation ability of GPU.
Xiuhong Li et al. design a new coordinated tiling and batching algorithm, which significantly improves the computational efficiency of small matrix in deep layers of CNNs \cite{gemm_coordinated}.
Andrew Anderson et al. propose two novel algorithms for convolution using GEMM that requires $O(MHW)$ and $O(KW+CH+CW)$ additional space respectively \cite{Low-Memory_gemm}. By reducing the space requirements of convolution, they improve data locality and parallel performance.
Fast Fourier Transform (FFT) is a computational tool commonly used for signal analysis. FFT-based methods compute convolutions as pointwise products in the Fourier domain while reusing the same transformed feature map many times \cite{b14}. The time complexity of FFT-based method is reduced compared with direct convolution. FFT-based methods usually perform better at a larger filter size.
Winograd is a method based on Winograd minimal filtering algorithms. Winograd method dramatically reduces the arithmetic complexity compared with direct convolution \cite{b15}. Winograd-based methods perform well for small kernels and small batch sizes. Although the dense methods which convert convolution to GEMM can better exploit the strong computing power of GPU, the large sparsity in the feature map is ignored. The methods proposed in this paper can skip the zero-value calculation in convolution, which will greatly reduce operations of multiplication and addition.

\subsubsection{FPGA or ASIC}
New architectures based on FPGA \cite{fpga_survey, embedded} or ASIC are usually more efficient than CPU or GPU platforms.
Srimat Chakradhar et al. propose the first CNN architecture to achieve real-time video stream processing \cite{firstCNNarch}. 
DianNao \cite{diannao} is an accelerator for large-scale CNNs, optimized on both performance and energy.
Manoj Alwani et al. construct a fused-layer CNN accelerator which fully exploits data reuse between adjacent CNN layers \cite{Fused-layer}. 
Eyeriss is an accelerator for state-of-the-art CNNs. It optimizes the energy efficiency of the entire system for various CNN shapes \cite{eyeriss}.
PRIME can accelerate CNN in ReRAM-based main memory with significant performance improvement and energy saving \cite{prime}.
FpgaConvNet is an end-to-end framework for the optimized mapping of CNNs on FPGAs \cite{fpgaConvNet}.
Tensor Processing Unit (TPU) is a custom ASIC with a MAC matrix multiply unit and a large software-managed on-chip memory \cite{tpu}.
Thinker is an energy efficient reconfigurable hybrid neural network processor fabricated in 65-nm technology \cite{thinker}.
Parana is a neural processor for hybrid neural network acceleration in consideration of thermal problem of 3D DRAM \cite{parana}.
PipeLayer is a ReRAM-based PIM accelerator for CNNs that supports both training and testing \cite{pipelayer}.
ShiDianNao is a CNN accelerator which is placed next to a CMOS or CCD sensor \cite{shidiannao}. 
Convolution Engine is specialized for the convolution-like data-flow that is common in computational photography, image processing, and video processing applications \cite{convolutionEngine}.
MAPLE is an accelerator for machine learning algorithms such as CNN. It has hundreds of simple PEs laid out in a two-dimensional grid \cite{maple}.
Maurice Peemen et al. propose a memory-centric accelerator for CNNs \cite{memory-centric}.
nn-X is a scalable and low-power coprocessor for real-time execution of CNNs \cite{nn-x}.
UniWiG \cite{UniWiG} is a unified architecture where GEMM can be accelerated using PEs. It can efficiently utilize FPGA hardware resources when computing all layers of CNN.
Sung-En Chang et al. implement a new architecture which applies different quantization schemes for different rows of the weight matrix for accelerating CNN on the FPGA platform \cite{MixandMatch}. 
Tensor Core \cite{tensorcore} in GPUs is good at accelerating CNNs by optimizing GEMM. Tensor Cores use mixed precision to compute GEMM, with only a minimal loss of precision in the output. 

\subsection{Sparse methods for accelerating CNNs}
The sparsity of CNN mainly comes from the commonly used RELU activation function and pruning. The sparsity of feature maps in deep network layers is usually more than 0.8, as shown in Fig. \ref{FigSparisity}. Therefore, exploiting sparsity has great potential to accelerate CNNs. The methods of accelerating CNNs by sparsity can also be classified into CPU/GPU and FPGA/ASIC. 

\subsubsection{CPU or GPU}
Although the deep layers in CNNs have high sparsity, this unstructured sparsity is difficult to exploit, so some algorithms have been developed to optimize sparse matrix multiplication (SpGEMM) \cite{cpu-sparseconv,SpMVGPU}.
Yuyao Niu et al. \cite{TileSpGEMM} propose a tiled parallel SpGEMM algorithm named TileSpGEMM. The main feature of TileSpGEMM is that the input and output sparse matrices are stored as multiple non-empty sparse blocks of the same size.
SparseRT is a code generator that leverages unstructured sparsity to accelerate sparse linear algebra operations in deep learning inference on GPUs \cite{sparseRT}.
Balanced Sparsity is a novel fine-grained sparsity approach to achieve high model accuracy with commercial hardware efficiently \cite{banlanced-sparsity}.
PCONV comprises a new sparsity dimention, fine-grained pruning patterns inside the coarse-grained
structures \cite{PCONV}.
Yue Zhao et al. effectively bridge the gap between deep learning and the special needs of the pillar HPC problem through a set of techniques on matrix representations, deep learning structure, and cross-architecture model migrations \cite{dl_spmv}.
Vijay Daultani et al. propose a new computation and memory efficient convolution algorithm for the inference phase, Sparse Direct Convolution (SDC) and a new representation for sparse filters, Compressed Sparse Offset (CSO) \cite{sdc-cso}.
Skip-Convolutions leverage a large number of redundancies in video streams and save computations \cite{skip-conv}.
Spartan is a lightweight hardware/software framework for accelerating DNN training on GPUs \cite{Spartan}, which can exploit activation sparsity detected during training. 
Zhuoran Song et al. propose a sensitivity-aware dropout method to achieve greater forward and backward training acceleration for CNN while reserving the accuracy \cite{gpu-friend}.
FalCon is a novel sparsity computing
scheme which can well adapt to the practical sparsity patterns while still maintaining efficient computing \cite{FalCon}.
Peter Ahrens et al. propose an automatic asymptotic scheduler for sparse tensor programs \cite{AutoSparse}.
In practice, current implementations of sparse convolution on GPU are usually based on sparse libraries such as cuSPARSE \cite{GPU-arXiv-2020, GPU-ICPP-2019,Lee2022}. The cuSPARSE library contains a set of basic linear algebra subroutines used for handling sparse matrices. When using cuSPARSE to calculate convolution, a three-step processing is performed on the feature maps and convolution kernels, namely extension, compression and sparse matrix computation. Previous sparse methods can hardly outperform cuBLAS-based convolution methods. Compared with the previous sparse methods, the sparse methods in this paper try to further reduce the data transfer not only between CPU and GPU but also between off-chip memory and on-chip memory of GPU.

\subsubsection{FPGA or ASIC}
New hardwares are designed to exploit sparsity in CNNs mostly based on FPGA or ASIC \cite{spots}. 
Earlier works \cite{Cnvlutin, Cambricon} only explore the sparsity of the feature maps, which is caused by the RELU activation function. These schemes take advantage of one-sided sparsity, with only zeros in the feature maps, while the filter values remain unchanged.
However, with the application of pruning, filters also exhibit sparsity. 
Gondimalla et al. \cite{SparTen} utilized zeros in both feature maps and filters to achieve more efficient sparse computation. 
Sparse Tensor Core \cite{STC} is an algorithm and hardware co-design, which is implemented on NVIDIA A100 GPU. 
PERMCNN is an energy-efficient hardware architecture for permuted diagonal structured CNNs \cite{PermCNN}.
TensorDash is a hardware-based technique that enables data-parallel MAC units to take advantage of sparsity in their input operand streams \cite{TensorDash}.
Sourya Dey et al. propose a reconfigurable hardware architecture to accelerate the training of CNN by sparsity \cite{b21}.
EIE is an energy efficient inference engine that performs inference on the compressed network model and accelerates the resulting sparse matrix-vector multiplication with weight sharing \cite{EIE}.
Scalpel customizes DNN pruning to the underlying hardware by matching the pruned network structure to the data-parallel hardware organization \cite{Scalpel}.
FuseKNA eliminates both ineffectual and repetitive additions in bitserial computation by exploiting bit repetition and bit sparsity in weights, for both convolutional and fully-connected layers \cite{FuseKNA}.
GoSPA is an energy-efficient high-performance globally optimized sparse CNN accelerator \cite{GoSPA}.
SIGMA is a flexible and scalable architecture that offers high utilization of all its PEs regardless of kernel shape and sparsity \cite{sigma}.
CNV uses hierarchical data-parallel units, allowing groups of lanes to proceed mostly independently enabling them to skip over the ineffectual computations \cite{Cnvlutin}.
Sparse ReRAM Engine exploits both weight and activation sparsity to eliminate ineffectual computation \cite{sparseReram}.


\section{Motivation} \label{motivation}
CNN extracts the characteristic information of input data through convolution layers. The convolution kernel performs multiplications and additions in the sample area throughout the feature map.
If the size of feature map is $i_{w}\times i_{h}$ and the kernel size is $k_{w}\times k_{h}$, a convolution operation requires $num_{mul}$ multiplications and $num_{add}$ additions, according to Eq. (\ref{equ:num_mul}) and Eq. (\ref{equ:num_add}) respectively.

\begin{equation}
	num_{mul}=(\frac{i_{w}-k_{w}}{c_{s}}+1)(\frac{i_{h}-k_{h}}{c_{s}}+1)(k_{w} k_{h})
	\label{equ:num_mul}
\end{equation}

\begin{equation}
	num_{add}= (\frac{i_{w}-k_{w}}{c_{s}}+1)(\frac{i_{h}-k_{h}}{c_{s}}+1)(k_{w} k_{h}-1)
	\label{equ:num_add}
\end{equation}

Feature maps of the deeper convolution layers are usually smaller, and the size can be as small as $5\times 5$. Convolution calculation on GPU is generally converted into GEMM, which is often more efficient for large matrix. Because the number of GPU threads is small due to the limited size of the feature map, GEMM cannot fully exploit GPU's computation ability.

\begin{figure}[t]
	\centering
	\includegraphics[width=\textwidth]{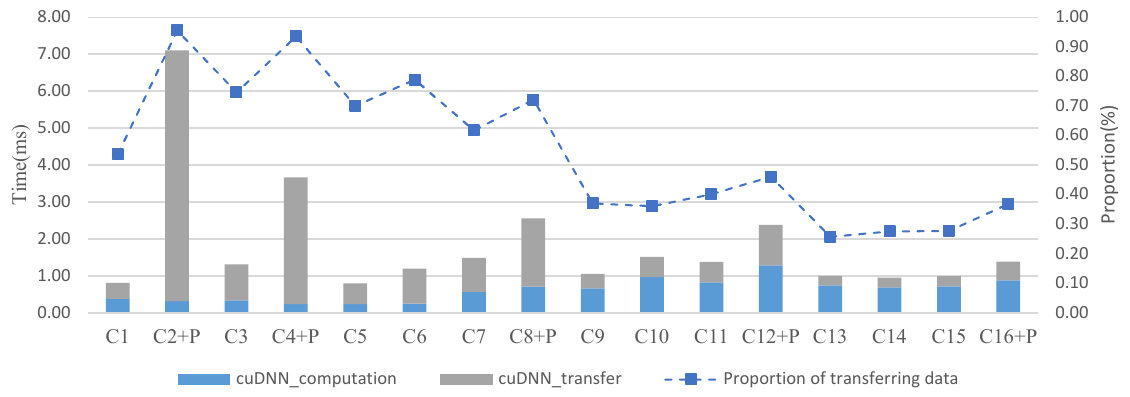}\\
\vspace{-5pt}
	\caption{Convolution (C) and pooling (P) of VGG-19 by cuDNN considering the data transfer between CPU and GPU. The total time is divided into data transfer time and computing time.  The left vertical coordinate is time (ms). The right vertical coordinate is proportion.}
	\label{breakdown_time}

\end{figure}

The sparsity of the deep feature map is relatively large due to RELU activation and pruning operations. We can see from Fig. \ref{FigSparisity} that the sparsity of the feature map which will enter the convolution layer can reach more than 0.7 for deep layers in the network \cite{b40}. When the feature map is extended as in Im2col, the matrix is even more sparse (grey bar). The traditional methods such as Im2col will calculate these zero values, which are useless for the final convolution result. 



Although there are some traditional compression formats for sparse matrices such as CSR \cite{CSR}, they cannot fully adapt to the characteristics of convolution operations. Sparse Tensor Cores (STC) can compute sparse matrix multiplication on GPUs \cite{STC}. However, STC uses an algorithm which prunes the matrix to a 2:4 sparse pattern. That means two values must be zero in each contiguous block of four values. Although such a method will improve the inference speed of CNNs, it will cause decrease of accuracy.

To address the above problems, we reduce the number of multiplications and additions in the convolution operation by compressing the feature map. In order to reduce the time consumption, we transform the traditional matrix conversion operation \cite{b12} into a compression operation which does not affect the accuracy of CNNs. After compression, the feature map is transformed to a storage format fully considering the characteristics of convolution operations, and the convolution calculation is transformed into sparse matrix-vector multiplication (SpMV) \cite{spmvxu}, which eliminates the calculation of redundant zero values. Furthermore, the process of data format transformation and SpMV only require one time global memory access.

On the other hand, traditionally, convolution layers are usually calculated in an independent GPU kernel. After the convolution results are transferred to CPU memory, the calculation of pooling layer is started, such as in Caffe \cite{Caffe} and PyTorch  \cite{Pytorch}. This will increase the traffic between CPU and GPU. 
As shown in Fig. \ref{breakdown_time},  the time for transferring data between CPU and GPU occupies a considerable proportion in the entire convolution and pooling calculation process especially for the first several layers. If the time consumption of data transfer between the CPU and GPU during convolution and pooling operations can be reduced, the total time of CNN can be reduced effectively. Furthermore, the time will be further reduced if the data transfer between global memory and shared memory is reduced on GPU.

\section{Sparse convolution method on GPU} \label{ecr}
In this section, we introduce a sparse convolution method for CNN on GPU. First, we present the whole procedure of CNN forward computing with the sparse convolution method. Then, a new storage format suitable for sparse convolution calculation on GPU is proposed. At last, convolution method based on the new storage format is described in detail.

\subsection{Whole procedure for CNN forward computing}
In order to better understand the proposed method, we design an algorithm procedure with only one-time  data  transfer  to  reduce off-chip memory traffic (traffic between CPU and GPU, and traffic between global memory and shared memory on GPU) for the CNN forward computing. The algorithm procedure is as follows. 

Step 0: Transfer input data (images/feature maps and filters) from CPU to GPU. Start GPU kernel.

Step 1: Load input data from global memory to shared memory and transform the data into the new format.

Step 2: Perform SpMV for convolution layer. One thread computes one convolution result.

Step 3: Perform pooling operations if there is a pooling layer after the convolution layer, and output the pooling result to global memory.

Step 4: Go to Step 1 unless it is the last pooling layer.

Step 5: Complete the computation for remaining layers.

Step 6: Transfer the result from GPU to CPU.

\subsection{Extended and compressed row storage format}

\begin{figure}[!htp]
	\centering
	\includegraphics[width=0.8\linewidth]{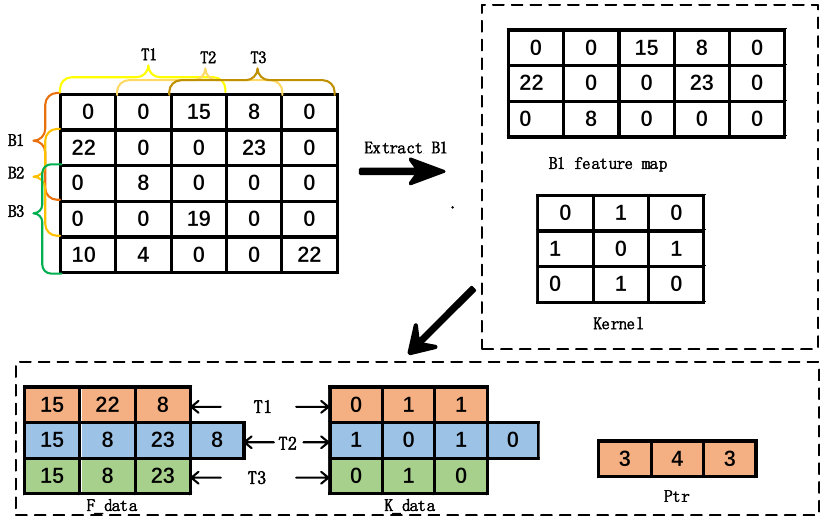}\\
	\caption{A novel storage format (ECR) for one feature map and one kernel. The size of feature map is $5\times 5$, kernel size is $3\times 3$, and the stride is $1$.}
	\label{ECR-format}
\vspace{-10pt}
\end{figure}

Zeros in feature maps for the convolution operation can lead to useless multiplications and additions. Therefore, we convert the feature map into a new storage format (Extended and Compressed Row, ECR), then the convolution can be converted to sparse matrix vector multiplication (SpMV).
As shown in Fig. \ref{ECR-format}, the feature map is divided into three convolution block rows ($B1$, $B2$ and $B3$) according to the size of the stride $c_s$ and the height of convolution kernel $k_{h}$. One convolution block row corresponds to one row of final convolution results. One thread block on GPU is assigned to compute one convolution block row. In each convolution block row, convolution kernel moves horizontally. Therefore, each convolution block row is divided into three convolution windows, and each convolution window corresponds to one convolution result. Each convolution result is computed by one thread ($T1$, $T2$ or $T3$) on GPU.

One block row $B1$ is extracted to describe the new storage format. We store the three non-zero values in  convolution window of $T1$ into $F\_data$ of $B1$ and store the corresponding values in the convolution kernel into $K\_data$  of $B1$. The non-zero values of all remaining convolution windows are stored into $F\_data$ in turn, along with the corresponding convolution kernel values into $K\_data$. Besides, the number of non-zero values in each convolution window is stored in $Ptr$, which is also the number of multiplications required for each thread. If there is no non-zero value for a convolution window, a value of $-1$ should be stored in $Ptr$ of $B1$ for markup. In all, non-zero values are stored in $F\_data$, kernel values are stored in $K\_data$, and the number of non-zeros values are stored in $Ptr$. Because the filter is also extended with feature map, bank conflicts will be reduced when all the threads visit the filter together in shared memory. 

\subsection{Load and transform data}

After images/feature maps and filters are transferred from CPU to GPU's global memory, GPU starts its work. The feature maps are loaded, extended and transformed into ECR format at the same time (Fig. \ref{ECR-format}). After that, the feature map and the filter in ECR format are stored in shared memory. 

\begin{figure}[h]
	\centering
	\includegraphics[width=0.7\linewidth]{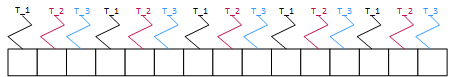}\\
	\caption{Coalesced global memory access by threads in a warp.}\label{thread}
\end{figure}

As shown in Fig. \ref{thread}, feature maps are stored in global memory as one-dimensional arrays for continuous access by adjacent threads. In this way, non-zero values of feature maps are loaded from global memory to shared memory, which can reduce the time of the global memory access. This advantage of coalesced global memory access performs best when the convolution stride is 1.

For a single feature map, we can allocate $\frac{i_{h}-k_{h}}{c_s}+1$ blocks, each of which contains $\frac{i_{w}-k_{w}}{c_s}+1$ threads. 
As shown in Algorithm \ref{formatalg}, 
the algorithm of one thread is described for converting its corresponding map and filter to ECR format.
$F_{data}$ and $K_{data}$ are declared shared memory. In Algorithm \ref{formatalg}, $temp$ is a counter for non-zero values in the feature map. Each thread needs to read $k_w{\times}k_h$ values of feature map from global memory (Line 2 and 3). Line 4 sets the address $offset$ that this thread needs to access in global memory. $offset$ equals $mapping(thread\_idx, data\_idx)$ which calculates the relationship between $thread\_idx$ in blocks and $data\_idx$ in global memory.
Line 5 judges the data value corresponding to the global memory position $offset$.
If it is a non-zero value, it is stored in $F_{data}$ with the position $thread\_idx*k_w*k_h+num_{nonzero}$, and the corresponding value in convolution kernel is stored in the corresponding position of $K_{data}$ (Line 6, 7 and 8).
After the loop, the number of non-zero values is stored in $Ptr$, and -1 is stored if there is no non-zero value (Line 12 to 16).


\begin{algorithm}[h]
	\caption{Storage format conversion algorithm for ECR}
	\label{formatalg}
	\begin{algorithmic}[1]
		
		\STATE $temp \gets 0$
		\FOR{$i=0$ to $k\_h$}
		\FOR{$j=0$ to $k\_w$}
        \STATE $offset \gets mapping(thread\_idx, data\_idx)$
		\IF{$input[offset]!=0$}
		\STATE $F\_data[thread\_idx*k\_w*k\_h+temp] \gets input[offset]$
		\STATE $K\_data[thread\_idx*k\_w*k\_h+temp] \gets kernel[i+j*k_w]$
		\STATE $temp++$
		\ENDIF
		
		\ENDFOR
		\ENDFOR
		\IF{$temp!=0$}
		\STATE $ptr[thread\_idx] \gets temp$
		\ELSE
		\STATE $ptr[thread\_idx] \gets -1$
		\ENDIF
		
	\end{algorithmic}
\end{algorithm}

\subsection{Perform SpMV for convolution}

\begin{figure}[!htp]
	\centering
	\includegraphics[width=0.8\linewidth]{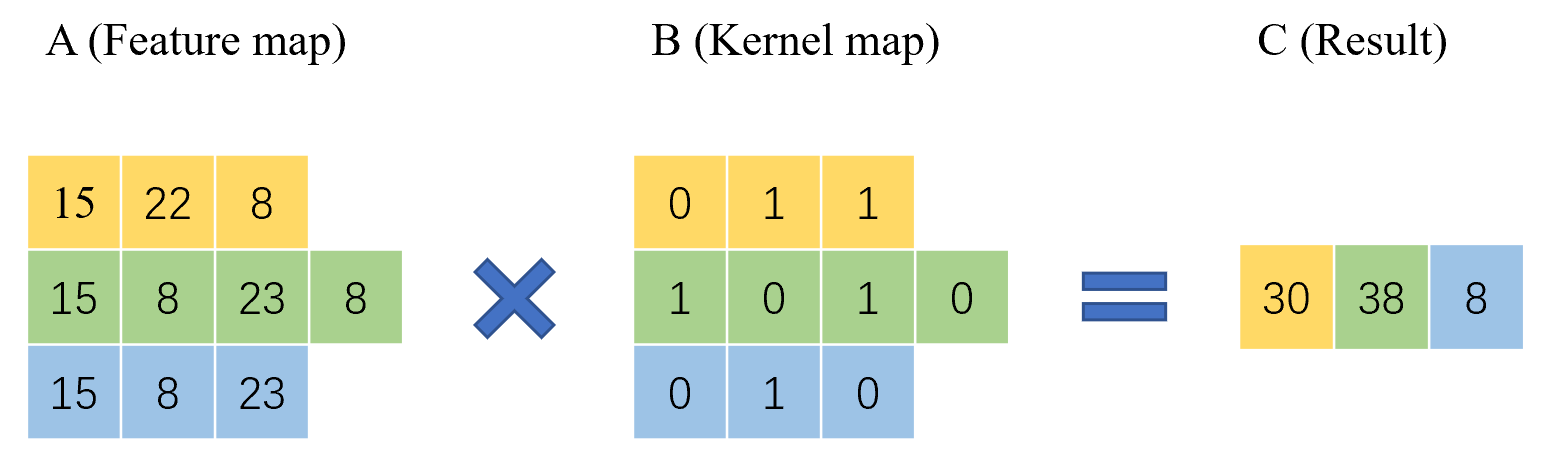}\\
	\caption{Sparse matrix vector multiplication. The values in the same color are calculated by the same thread.}
	\label{mul}
\end{figure}

In this subsection, we will describe the algorithm for sparse convolution calculation based on ECR and use Fig. \ref{mul} as an example to analyze how the number of multiplication and addition is reduced.

In the previous subsection, we obtained a new data format ECR, $F_{data}$ for feature map, $K_{data}$ for kernel data, and $Ptr$ for the number of non-zero values. As shown in Fig. \ref{mul}, we can consider $F_{data}$ as a sparse matrix, and $K_{data}$ as a vector with varying lengths. The length of the vector is the same as the corresponding value in $Ptr$. Therefore, the convolution operation is converted to a variant of SpMV, and zero values in the feature map are skipped without affecting convolution result.

As shown in Fig. \ref{mul}, for GPU implementation, one thread multiplies a row in the compressed feature map (A) with the corresponding vector in the kernel map (B) to obtain a value in a convolution result matrix (C). Therefore, 
a thread block contains $\frac{i_{w}-k_{w}}{c_s}+1$  threads, and can get one row of the final convolution results.
After parallel calculation by multiple thread blocks, the convolution result for the whole feature map can be obtained. 

The pseudo-code for a single thread to get a convolution result is described in Algorithm \ref{convalg}. Line 1 accesses the corresponding $ptr$ vector in the shared memory to determine whether there are non-zero values stored in $F_{data}$. If $ptr[thread\_idx]$ is $-1$, it is immediately judged that no operation is needed, and the output is 0 (Line 2). If the value stored in $ptr[thread\_idx]$ is not $-1$, the values in $K_{data}$ and $F_{data}$ are accessed in turn and multiply-add operations are performed (Line 4$\sim$6). The convolution result is finally obtained and stored into global memory for the computation of next layer (e.g. pooling layer).

As shown in Fig. \ref{ECR-format} and Fig. \ref{mul}, for three convolution block rows (B1, B2, and B3), the conventional algorithm requires 24 additions and 27 multiplications, while our algorithm only requires 7 additions and 10 multiplications, reducing 71\% additions and 63\% multiplications. Besides, $K_{data}$, $F_{data}$, and $ptr$ are all stored in shared memory. Therefore, the proposed method can greatly improve the speed of convolution and accelerate the whole CNN.

\begin{algorithm}[h]
	\caption{Convolution algorithm for ECR}
	\label{convalg}
	\begin{algorithmic}[1]
		\IF{$ptr[thread\_idx] ==-1 $}
		\STATE $output[thread\_idx+block\_idx*blockDim.x] \gets 0$
		\ELSE
		\FOR{$i=thread\_idx*k_w*k_h$ to $thread\_idx*k_w*k_h+ptr[thread\_idx]-1$}
		\STATE$output[thread\_idx+block\_idx*blockDim.x] += F_{data}[i]*K_{data}[i] $
		\ENDFOR
		\ENDIF
	\end{algorithmic}
	
\end{algorithm}

\subsection{Other steps for CNN forward computing}

The convolution results are processed by RELU before they are outputted to the global memory. As threads in different thread blocks cannot communicate through shared memory, we have to store the activation results into global memory before pooling operation starts. Because pooling operation itself take up only a little time \cite{b39}, we don't discuss it's implementation here.
After pooling results are outputted to global memory, a new round is started by loading the new feature map and filter to shared memory (Fig. \ref{thread}), and transforming them into ECR format  for the next convolution layer. This process is repeated until all convolution layers and pooling layers are finished.

\section{Combining convolution and pooling} \label{pecr}

Traditionally, after the convolution results are transferred from GPU to CPU, the calculation of pooling layer is started. This will increase the cost of data transfer. In this section, we propose an optimization method which not only exploits the sparsity of feature map but also calculates convolution and pooling together without transferring intermediate results. 

\subsection{Whole procedure  for CNN forward computing}
For the forward computing of CNN, the algorithm procedure should also try to ensure one-time data transfer from CPU to GPU. Therefore, the traffic between CPU and GPU is reduced. For better understanding the proposed method, we present the whole process of CNN forward computing as follows.

Step 0: Transfer the images and filters from CPU to GPU. Start GPU kernel.

Step 1: Load data (images/feature maps and filters) from global memory to shared memory. At the same time, transform the data to the new format.

Step 2: Perform SpMV for the convolution layer. One thread computes one convolution result.

Step 3: Several convolution results, e.g. 4 (2$\times$2), are used to get a pooling result, which is stored to global memory. 

Step 4: Go to Step 1 unless it is the last pooling layer.

Step 5: Complete the computation for remaining layers.

Step 6: Transfer the final result from GPU to CPU.

\subsection{Sparse storage format considering both convolution and pooling}

In the traditional convolutional neural network, the result of the convolution operation will enter the pooling layer after the activation operation. If the convolution results are transferred to the CPU side, and then again to the GPU side for pooling operations, it will increase the overall CNN calculation time due to the bandwidth limitation between GPU and CPU. Therefore, we redesign the calculation process and combine the pooling operation with the convolution calculation by proposing a new storage format which can exploit the sparsity of the feature map. 

\begin{figure}[h]
	\centering
	\includegraphics[width=0.8\linewidth]{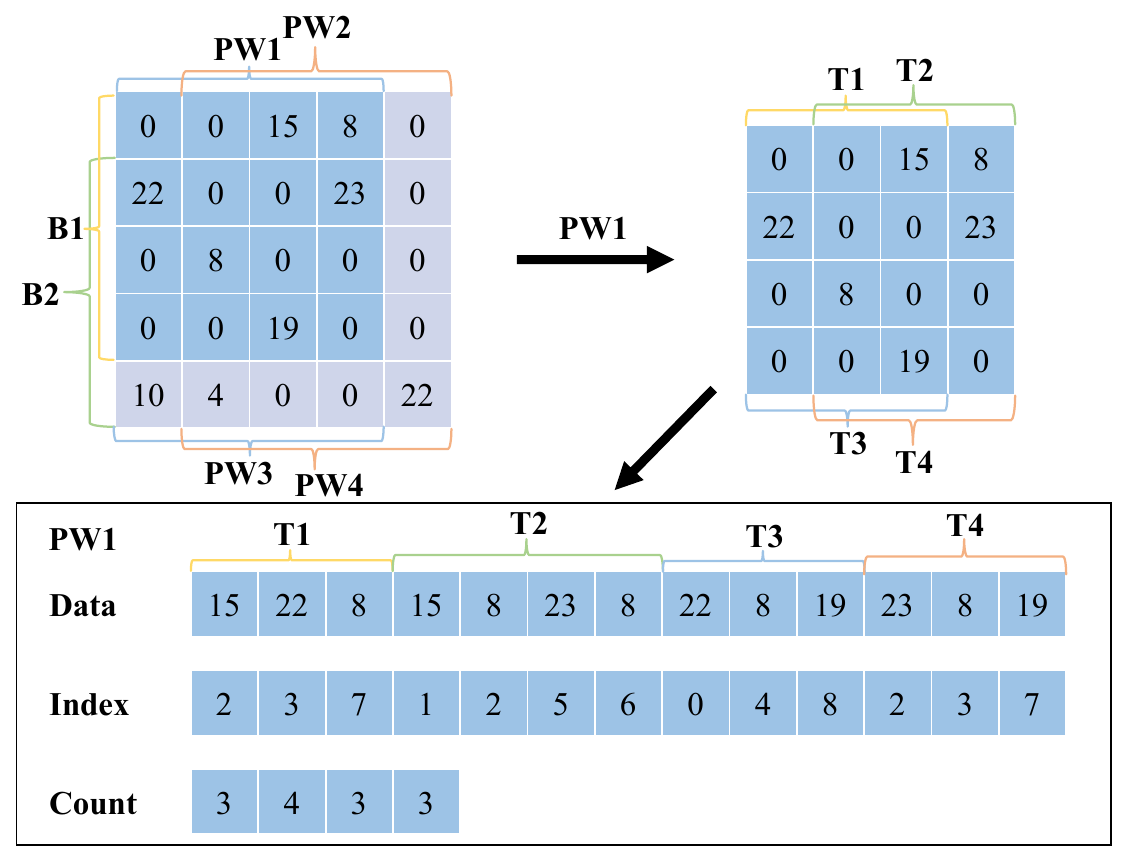}
	\caption{PECR format. The size of feature map is $5\times 5$. The size of convolution kernel is $3\times 3$. The size of pooling window (PW1, PW2, PW3, and PW4) after convolution is $2\times2$. The stride of both convolution and pooling is $1$.}\label{PECR}
\end{figure}

The new storage format is named as Pooling-pack Extended and Compressed Row (PECR) format. As shown in Fig. \ref{PECR}, the feature map is horizontally divided into two large pooling-pack rows ($B1$ and $B2$) according to the size of the stride $c_s$ and the height of convolution kernel $k_{h}$. One pooling-pack row corresponds to one row of pooling results. One thread block on GPU is assigned to compute one pooling-pack row. Each pooling-pack row is divided into two pooling windows, and each pooling window corresponds to one pooling result. 
Each pooling result is obtained by four threads ($T1$, $T2$, $T3$, and $T4$ ) on GPU. 

Non-zero values of all convolution windows in the corresponding pooling window are stored into  $Data$ in an order of left-to-right and top-to-bottom.
In order to complete the convolution calculation, the index pointing to corresponding value in the filter is stored in $Index$. 
Using such a storage method, each thread will process a sub-feature map with the size of $k_w\times k_h$ in the original feature map. $k_w\times k_h$ is the size of convolution kernel. 
A pooling window with the size of $T_w\times T_h$ will get a pooling result. $T_w$ equals $k_w+c_s(p_w-1)$, and $T_h$ generally equals $T_w$. Each thread in the pooling window will process one convolution window. For example, there are $2\times2$ pooling windows , PW1, PW2, PW3, and PW4, and there are $2\times2$ threads in a pooling window, T1, T2, T3, and T4 in Fig. \ref{PECR}.
%


\begin{algorithm}[t]
	\caption{Convert feature map and filter into PECR format}
	\label{alg-covert-pecr}
	\begin{algorithmic}[1]
		\STATE $pos \gets thread\_id * k_w * k_h$
		\STATE $cnt \gets 0$
		\STATE $num \gets 0$
		\FOR {$i=0$ to $k_h$}
		\FOR {$j=0$ to $k_w$}
		\STATE $ offset \gets mapping(thread\_id, data\_id)$
		\IF {$Input[offset] != 0$}
		\STATE  $Data[pos+cnt] \gets Input[offset]$
		\STATE $Index[pos+cnt] \gets i*k_h+j$
		\STATE $cnt += 1$
		\STATE $num += 1$
		\ENDIF
		\ENDFOR
		\ENDFOR
		\STATE $count[threadId\_block] \gets num$
	\end{algorithmic}
\end{algorithm}

\subsection{Load and transform data}

The feature maps and filters are loaded from global memory to shared memory in a coalesced way similar to Fig. \ref{thread}. At the same time, the input data are transformed to PECR format. 
The process of loading and converting data to PECR format for one GPU thread is described in Algorithm \ref{alg-covert-pecr}. Each thread processes one convolution window. For each convolution window ($k_w*k_h$) in the feature map, the thread loads non-zero values from global memory to shared memory, and computes the according index in the filter. Therefore, the time  complexity of Algorithm \ref{alg-covert-pecr} is $O(k_w*k_h)$. Each feature map is processed by $p_w*p_h*{n_o}^2$ threads, and $n_o$ is explained by Eq. (\ref{equ:n_o}). 

\begin{equation}
	n_o = \frac{(I_w-k_w+c_s-c_s*p_w+p_s*c_s)}{p_s*c_s}
	\label{equ:n_o}
\end{equation}


\subsection{Convolution and pooling}

\begin{figure}[!htp]
	\centering
	\includegraphics[scale=0.5]{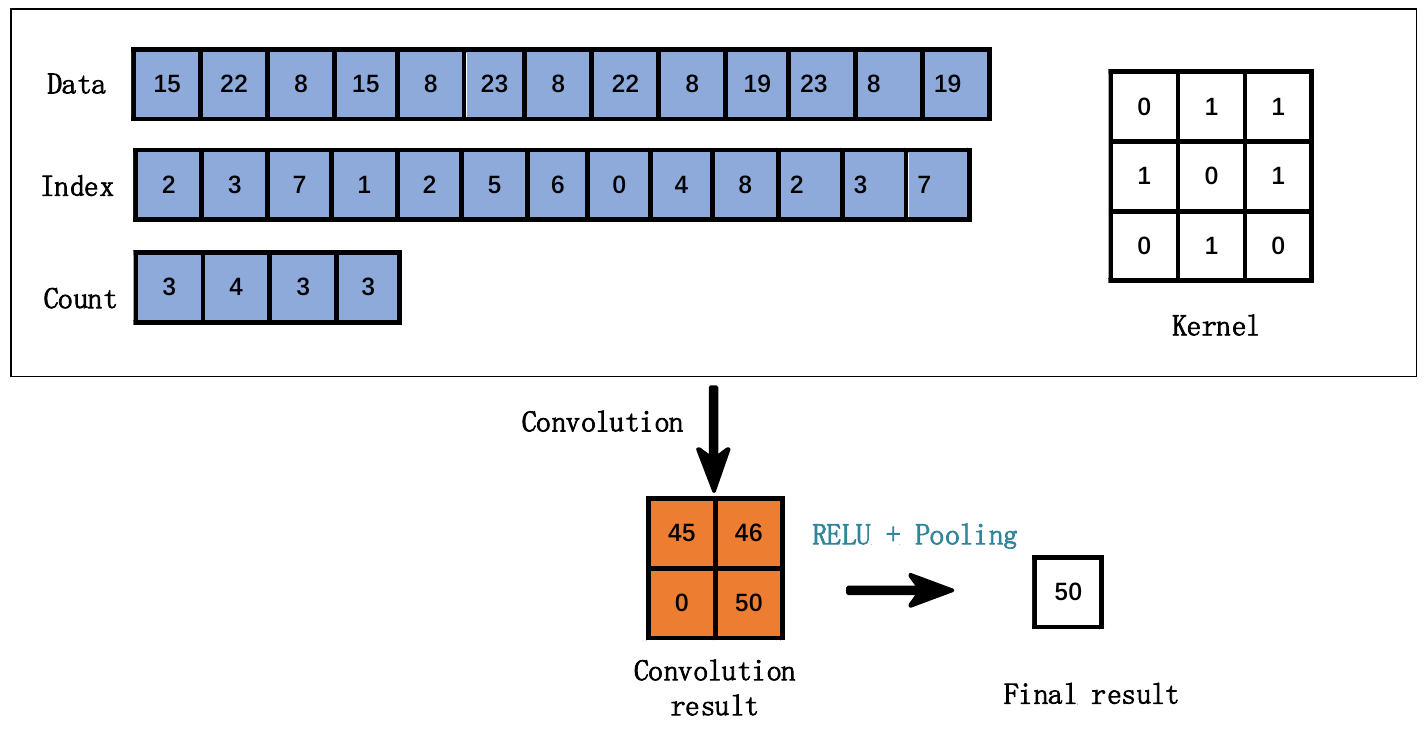}\\

	\caption{Convolution and pooling with PECR format.}\label{cp1}
\vspace{-5pt}
\end{figure}

After the feature map and the filter are transformed to PECR format and stored in shared memory, the convolution and pooling operations are started. 
Based on PECR format, each thread performs one SpMV operation to get one convolution result (Fig. \ref{cp1}). Four threads in the same warp get four convolution results in one pooling window. After that, RELU is used as the activation function, and a value less than zero is set to zero. Then, maximum pooling is used to get the final pooling result. 
By combining convolution and pooling, pooling result is obtained without data transfer between shared memory and global memory. %

Algorithm \ref{alg-conv-pool-pecr} describes the convolution and pooling algorithm based on PECR. A thread needs to process one convolution window. $p_w*p_h$ threads compute one pooling result by reduce operation (Line 9$\sim$13). 
Because different thread blocks cannot share data in shared memory, the pooling results are outputted to global memory. 
However, if only one thread block is started, the pooling results can be outputted to shared memory for the computation of the next convolution layer.

\begin{algorithm}[h]
	\caption{Convolution and pooling based on PECR}
	\label{alg-conv-pool-pecr}
	\begin{algorithmic}[1]
	    \STATE $tid \gets threadId\_block$
	    \STATE $temp \gets 0$
	    \STATE $pos\_out \gets thread.y+blockId*blockDim.y$
	    \STATE $reduce\_id \gets reduce\_stride + threadIdx.y * blockDim.x$
		\FOR {$i=pos$ to $pos+count[thredId\_block]$}
		\STATE $temp += data[i]*kernel[Index[i]]$
		\ENDFOR
		\STATE $conv[tid]=temp$
		\FOR {$restride = blockDim.x / 2$ to $0$ by $restride>>1$}
		\IF{$tid<reduce\_id$}
		\STATE $conv[tid] \gets max(conv[tid],conv[tid+restride])$
		\ENDIF
	    \ENDFOR
	    \STATE $output[pos\_out] \gets conv[threadIdx.y * blockDim.x]$
	\end{algorithmic}
\end{algorithm}

\vspace{-15pt}
\section{Discussion} \label{discussion}

Some CNN-based models have only a few pooling layers after convolution layers such as ResNet, and some models have more pooling layers such as VGG. Therefore, we can use the ECR method for convolution layers without a following pooling layer and use the PECR method for convolution layers with  a following pooling layer when we implement the proposed methods for an entire deep neural network model. The above strategy is named as Optional Convolution and Pooling Algorithms (OCPA).

We describe in detail the sparse convolution algorithm for one feature map and one kernel. The proposed ECR method can complete extension, compression, and sparse matrix computation by only accessing global memory once, which can effectively reduce the off-chip memory traffic. However, in the actual application, multiple feature maps and multiple kernels can be included in one convolution layer. Therefore, we implement different batch sizes in the experiment. 
The proposed algorithm can be extended to process this case by increasing the number of GPU threads. Since the amount of calculation in a single thread is reduced by the proposed algorithm, the calculation speed can also be improved with more threads. In addition, similar to Sparse
Tensor Core in the Ampere architecture \cite{b38}, the proposed method can also be applied to the hardware design. 

We present the convolution and pooling algorithm with PECR for one feature map and one kernel. However, there are sometimes multiple feature maps and multiple filters. In this case, more traffic between CPU and GPU is needed for cuDNN. Therefore, time consumption of data transfer between CPU and GPU can be further reduced with the proposed method, and the speedup will also be improved. 
It is worth noting that in the multi-channel convolution calculation, the data of other channels follow the same operation in turn. After all the channels are compressed, SpMV starts to run to ensure that the calculation results are correct.

%

\begin{table}[htp]
\vspace{-5pt}
	\caption{Experiment environment}\label{computer}
	\centering
	\resizebox{0.8\linewidth}{!}{
	\begin{tabular}{|p{0.2\linewidth}|p{0.7\linewidth}|}
		\hline
		CPU & Intel Xeon Gold 6132 CPU, 128GB DRAM \\
		\hline
		GPU & NVIDIA GeForce GTX 2080Ti  \\
		\hline
		OS & Ubuntu18.04 \\
		\hline
		CUDA & 10.2 \\
		\hline
		cuDNN & 7.6.5 \\
		\hline
	\end{tabular}}
\vspace{-10pt}
\end{table}


\begin{table}[h]
	\centering
	\caption{Speedup of ECR over cuDNN (FAST) and cuSPARSE for a single convolution layer}\label{table}
	\resizebox{0.8\linewidth}{!}{
	\begin{tabular}{|c|c|c|c|c|c|}
		\hline
		Network     & Layer         & Size         & Sparsity &	Speedup(cuDNN)  &  Speedup(cuSPARSE)\\\hline
		LeNet       & Conv2         & 11$\times$11 & 0.95     & 2.78            &  3.35             \\\hline
		AlexNetC	& Conv3	        & 6$\times$6   & 0.9      & 2.29            &  3.54             \\\hline
		AlexNetI	& Conv4	        & 5$\times$5   & 0.9      & 2.61	        &  3.66             \\\hline
		GoogLeNet	& Inception4a.1 & 14$\times$14 & 0.9      & 2.24            &  3.36             \\\hline
		GoogLeNet	& Inception4a.2 & 14$\times$14 & 0.9      &	2.52            &  2.97             \\\hline
		GoogLeNet	& Inception4e.3 & 14$\times$14 & 0.9      &	2.24            &  3.34             \\\hline
		GoogLeNet	& Inception5a.1 & 7$\times$7   & 0.95     &	2.74            &  3.79             \\\hline
		GoogLeNet	& Inception5a.2 & 7$\times$7   & 0.9      &	2.56            &  3.65             \\\hline
		GoogLeNet	& Inception5b.3 & 7$\times$7   & 0.95     &	2.73            &  3.00             \\\hline
		GoogLeNet	& Inception4a.7 & 7$\times$7   & 0.95     & 2.69            &  3.99             \\\hline
		VGG-19     	& conv13        & 14$\times$14 & 0.85     & 2.34            &  3.62             \\\hline
		VGG-19      & conv15        & 14$\times$14 & 0.83     & 2.47            &  3.75             \\\hline
		ResNet-50   & conv42        & 14$\times$14 & 0.86     & 2.54            &  3.82             \\\hline
		ResNet-50   & conv46        & 7$\times$7   & 0.9      & 2.45            &  3.53             \\\hline
        \end{tabular}}
\end{table}

\section{Experiments} \label{experiments}

In this section, we present and analyze the experimental results of the proposed algorithms with various neural network models on the GPU platform (Table \ref{computer}). The code for this paper has been open source\footnote{https://github.com/sunnchioo/OCPA}.
%


%
		
%
%
		
%

\begin{figure}[H]
    \centering
    \subfigure[VGG-19]{
		\begin{minipage}[t]{0.48\linewidth}
		\centering
        \includegraphics[width=\textwidth, height=0.4\textwidth]{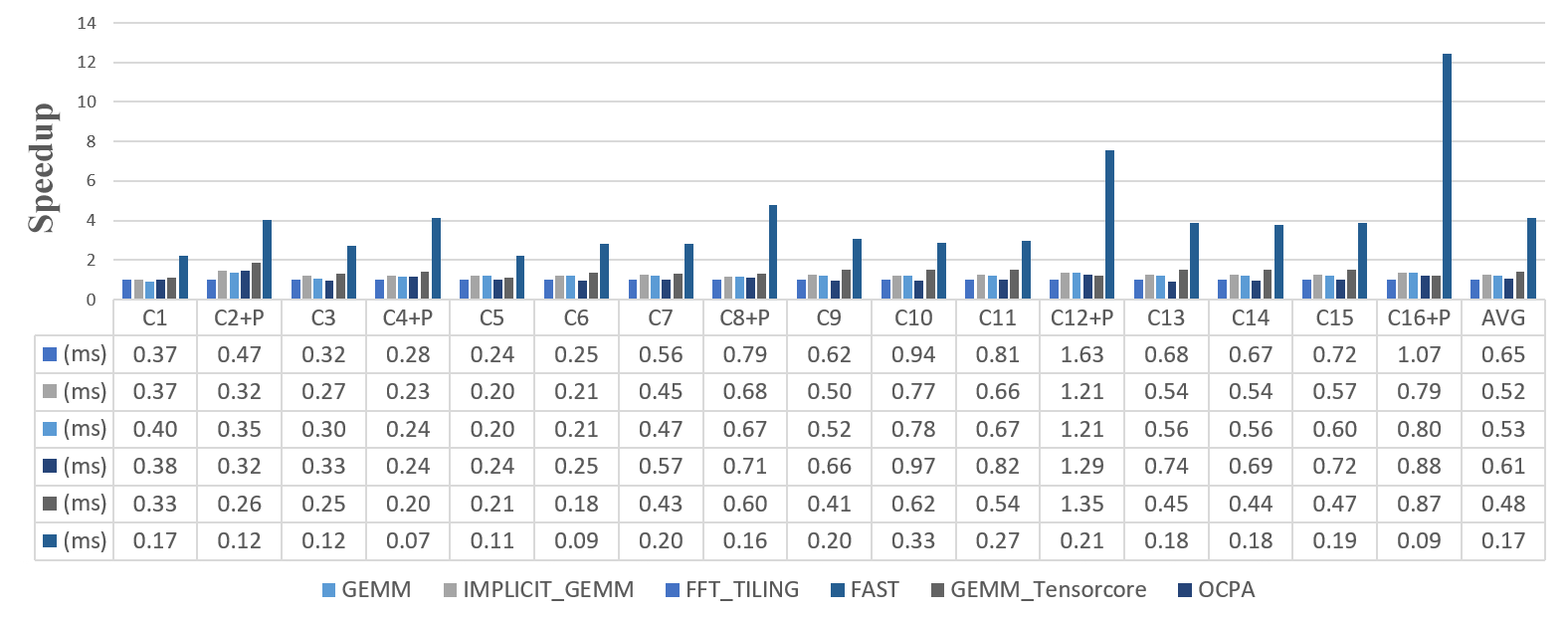}
		\label{speedup-of-vgg19}
		\end{minipage}
    }
    \subfigure[ResNet-50]{
        \begin{minipage}[t]{0.48\linewidth}
		\centering
        \includegraphics[width=\textwidth, height=0.4\textwidth]{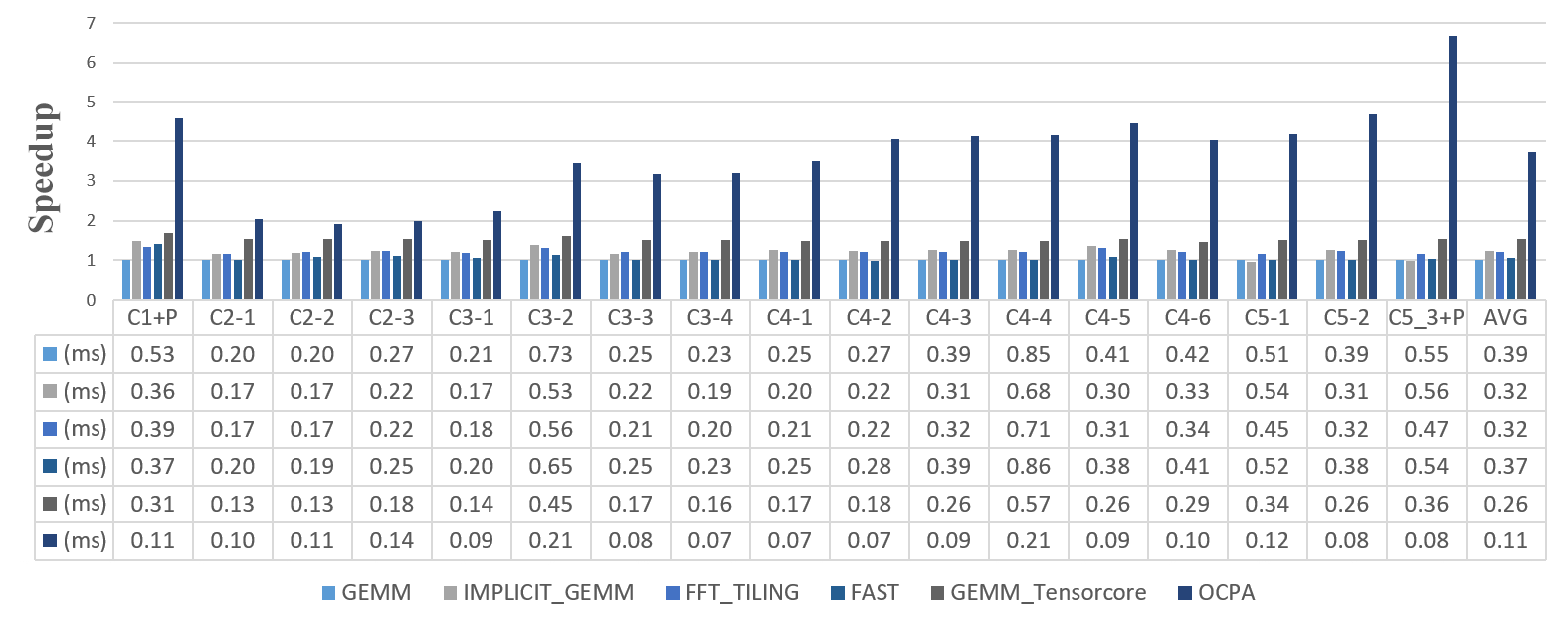}
		\label{speedup-of-resnet50}
		\end{minipage}
    }
 
    \subfigure[DenseNet-121]{
		\begin{minipage}[t]{0.48\linewidth}
		\centering
		\includegraphics[width=\textwidth, height=0.4\textwidth]{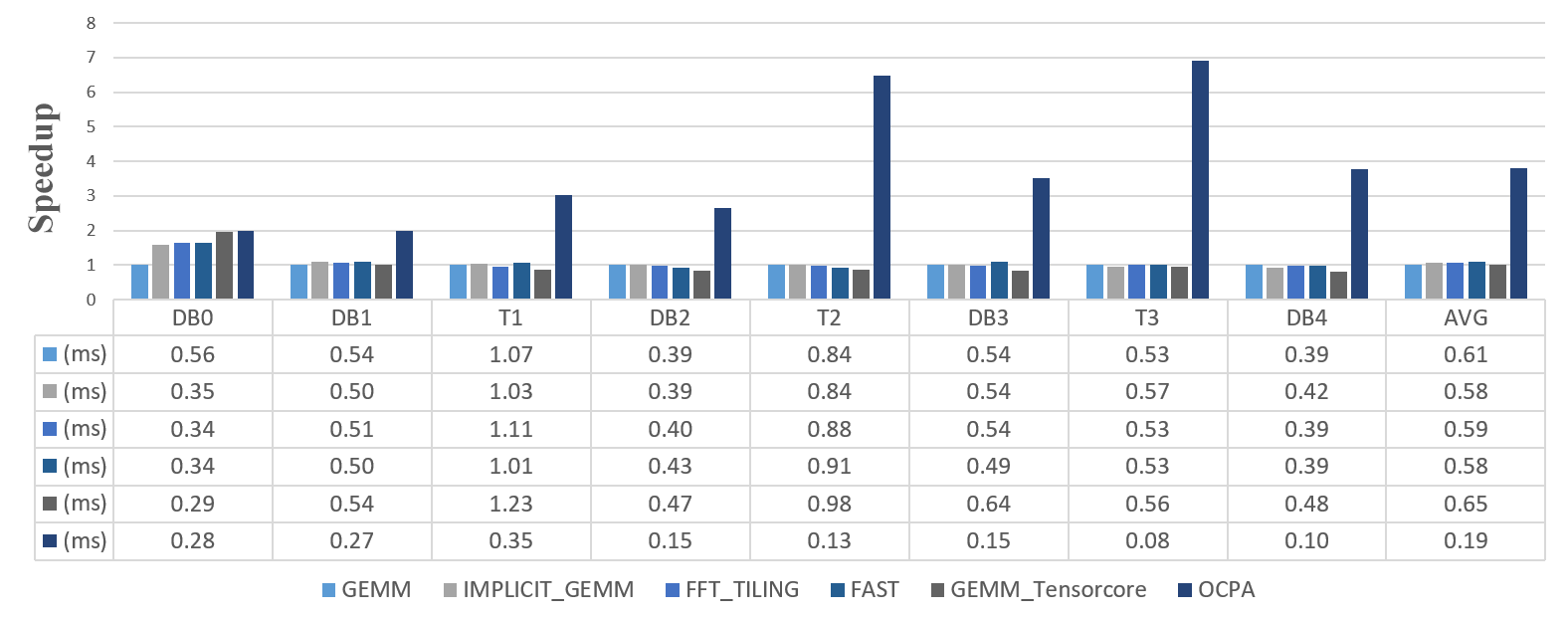}
		\label{speedup-of-densenet}
		\end{minipage}
  }
  \subfigure[RegNetX-16GF]{
		\begin{minipage}[t]{0.48\linewidth}
		\centering
	\includegraphics[width=\textwidth, height=0.4\textwidth]{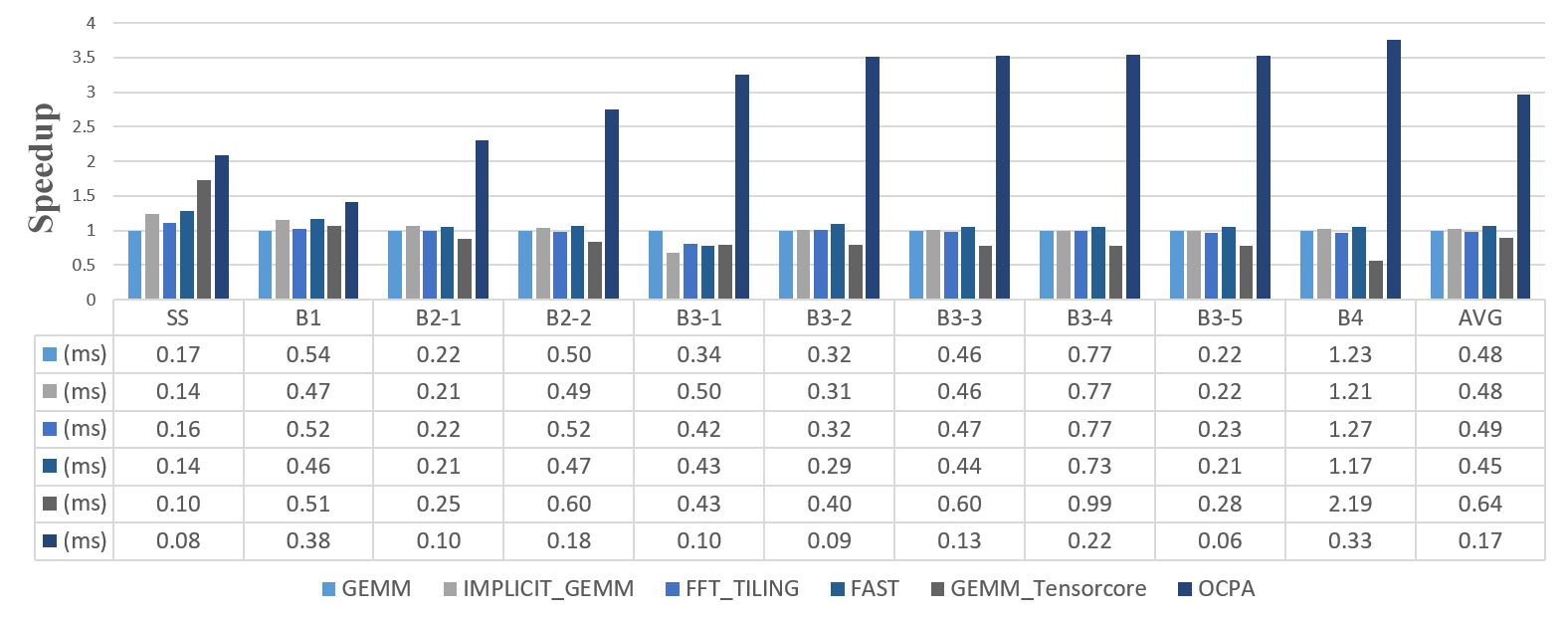}
		\label{speedup-of-regnet}
		\end{minipage}
  }
  \centering
  
\vspace{-5pt}
    \caption{Speed comparison between OCPA and other methods for VGG-19, ResNet-50, DenseNet-121, and RegNetX-16GF. The x-axis represents layers of neural networks. The y-axis represents the acceleration effect of different methods over GEMM.}
    \label{Speedup_over_gemm}
\vspace{-10pt}
\end{figure}

	
		

\subsection{Speed comparison for convolution layers}



In this subsection, we carried out speed comparison experiments for a single convolution layer. 
As shown in Table \ref{table}, the ECR method is used to process convolution layers from different models, such as LeNet, AlexNet, GoogLeNet, VGG-19 and ResNet-50.

We can see that the speedups for convolution layers are more than 2.2× compared with cuDNN (FAST), which can automatically choose a best implementation in cuDNN.
The reasons are mainly in two aspects. (1) For feature maps in deep networks, the size is very small comparing with the initial input feature map.
The traditional GEMM-based method in cuDNN is not suitable for matrix multiplication with small feature maps \cite{b15}. The ECR method reduces the amount of computation for a single thread according to the characteristic of sparsity and performs better for these small feature maps. (2) When using cuDNN to calculate convolution layers, it first uses Im2col to extend the feature map, and then convert the convolution calculation into matrix multiplication. This increases the number of global memory accesses. By comparison, ECR reads the values used for each convolution result from global memory only once.

Furthermore, ECR achieves up to 3.99× speedup for a single convolution layer compared with the cuSPARSE-based method.  
Although cuSPARSE-based method can skip the computation of zero values, it does not reduce the time for global memory accesses.
For the cuSPARSE-based method, the feature map is first extended as in Im2col, and then converted to the CSR format which is inputted to cuSPARSE for the final convolution result. These operations are separated, causing the program to repeatedly read data in global memory.
Note that the time of data-format conversion is included in the performance measurement of the ECR method.

\vspace{-5pt}

\subsection{Speedup for VGG, ResNet, DenseNet and RegNet}

Some convolution layers don’t have a following pooling layer in some CNNs, so we can use OCPA to implement the entire network, and compare OCPA with cuDNN and cuSPARSE-based method.

\subsubsection{Compared with cuDNN}

We implement the entire network using four cuDNN methods (GEMM, IMPLICIT\_GEMM, FFT\_TILING and FAST), Tensor Core-based method, and OCPA.
Fig.\ref{speedup-of-vgg19} shows the time and the speedup of other methods over the GEMM method in cuDNN for VGG-19. There are 16 convolution layers in VGG-19, 5 of which have a following pooling layer. We use C+P to present a convolution layer with a pooling layer which is computed by the PECR method. For a single convolution layer, the ECR method is used. Note that the time of data-format conversion is also included in the performance measurement.
We can see that OCPA achieves better performance than four cuDNN methods. 
On average, the speedup of OCPA over the baseline method cuDNN (GEMM) is $4.14\times$, and the speedup over cuDNN (FAST) is $3.91\times$ for convolution layers and convoluton+pooling layers in the VGG-19 network. In addition, the speedup over GEMM (Tensor Core) is $2.97\times$.
The speedup of OCPA over cuDNN (FAST) is $1.97\times$ for the entire VGG-19.

Fig. \ref{speedup-of-resnet50} shows the time and the speedup of different methods over GEMM in cuDNN for ResNet-50. 
We can see that OCPA achieves the best performance. On average, the speedup of OCPA over cuDNN (GEMM) is $3.73\times$, and the speedup over cuDNN (FAST) is $3.53\times$ for convolution layers and convoluton+pooling layers in the ResNet-50 network. Furthermore, the speedup over GEMM (Tensor Core) is $2.44\times$.
At last, we get the speedup of OCPA over cuDNN (FAST) is $2.23\times$ for the entire ResNet-50.
Because VGG-19 has more convolution+pooling layers than ResNet-50, the average speedup of VGG-19 for convolution layers and convolution+pooling layers is larger than that of ResNet-50.

Fig.\ref{speedup-of-densenet} shows the speedup of other methods over the GEMM method in cuDNN and the computation time of each layer for DenseNet-121. There are 120 convolution layers in DenseNet-121, 3 of which have a following pooling layer in Transition layers (T). Several convolution layers are combined to form a DenseBlock (DB), so we use the average time of the convolution layers in one DB to stand for the time of this DB. Using OCPA to implement the entire network, we can see that it has better performance than the other five methods.  On average, the speedup of OCPA over cuDNN (GEMM) is 3.79×, and the speedup over cuDNN (FAST) is 3.48× for convolution layers and convoluton+pooling layers in DenseNet-121. Since the feature maps of the first few convolutional layers of the model are less sparse, the speedup is smaller than that of the subsequent convolutional layers. Furthermore, the speedup over GEMM (Tensor Core) is 3.74×, and the speedup of OCPA over cuDNN (FAST) is 2.74× for the entire DenseNet-121. 

Fig. \ref{speedup-of-regnet} shows the speedup and the computation time of each layer using different methods for RegNetX-16GF which has 71 convolution layers in total. Several convolution layers are combined to form a Block (B), so we use the average time of the convolution layers in one Block to stand for the time of this Block. SimpleStemIN (SS) is the first part of RegNetX-16GF which contains a convolution layer. Since this model does not have a pooling layer immediately following the convolutional layer, we only use ECR in OCPA. On average, the speedup of OCPA over cuDNN (GEMM) is 2.96×, and the speedup over cuDNN (FAST) is 2.79× for convolution layers. Furthermore, the speedup over GEMM (Tensor Core) is 3.30×. At last, we get the speedup of 1.58× over cuDNN (FAST) for the entire RegNetX-16GF. 

\subsubsection{Compared with cuSPARSE}
\begin{figure}[!h]
		\centering
	\subfigure[VGG-19]{
		\begin{minipage}[t]{0.48\linewidth}
			\centering
			\includegraphics[width=\textwidth]{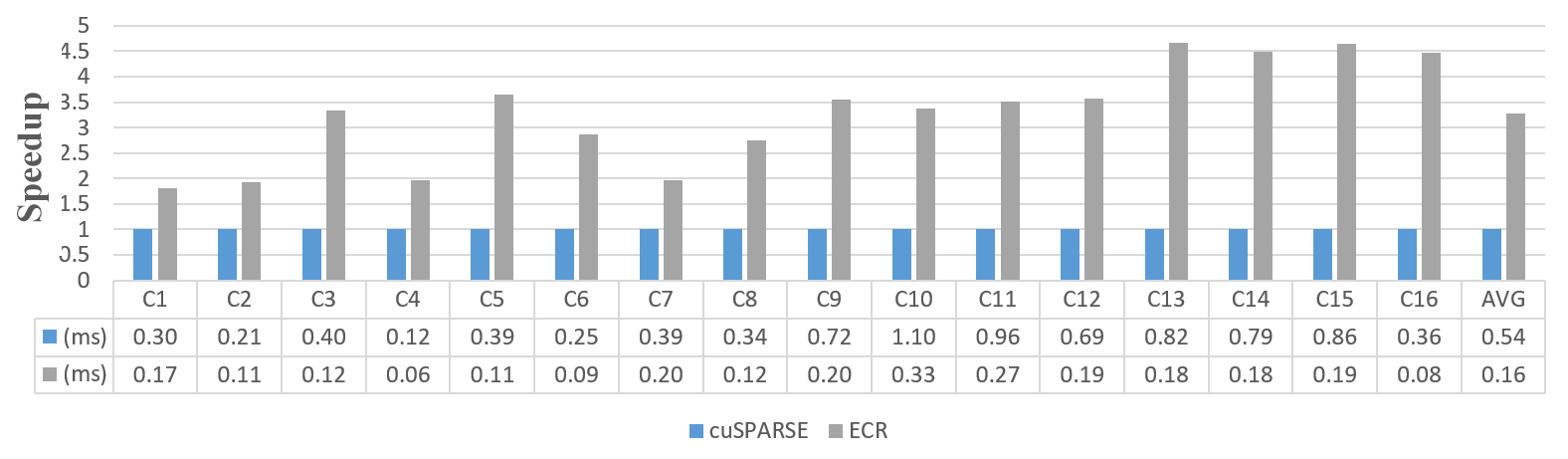}
			\label{cusparse_vgg}	
		\end{minipage}
	}
	\subfigure[ResNet-50]{
		\begin{minipage}[t]{0.48\linewidth}
			\centering
			\includegraphics[width=\textwidth]{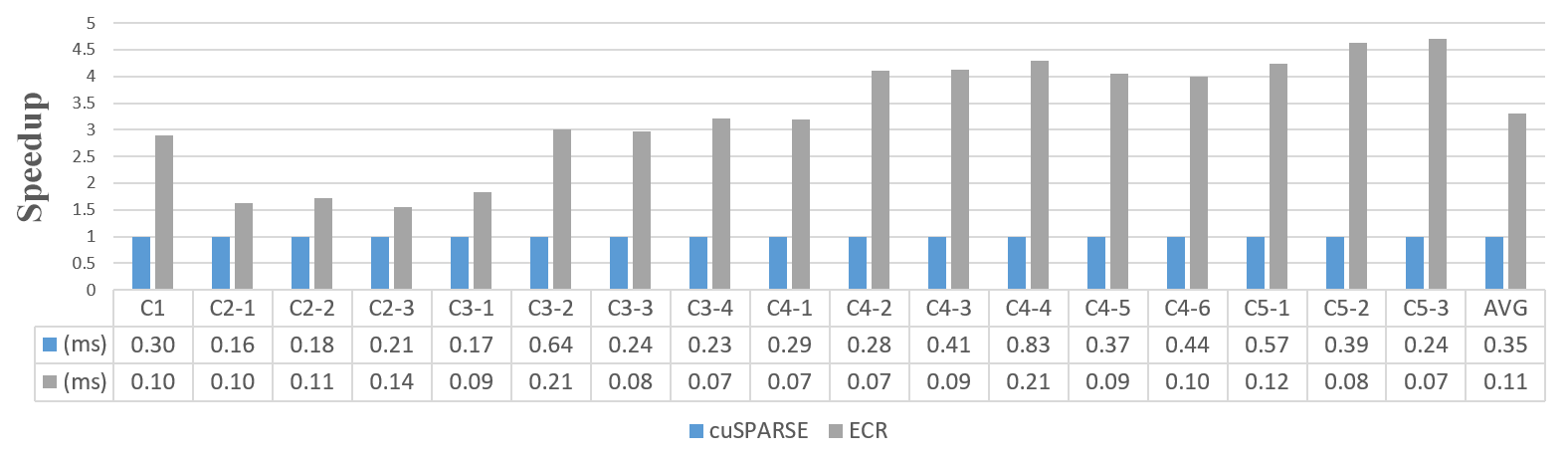}
			\label{cusparse_resnet}
		\end{minipage}
	}
 
 	\subfigure[DenseNet-121]{
		\begin{minipage}[t]{0.48\linewidth}
			\centering
			\includegraphics[width=\textwidth]{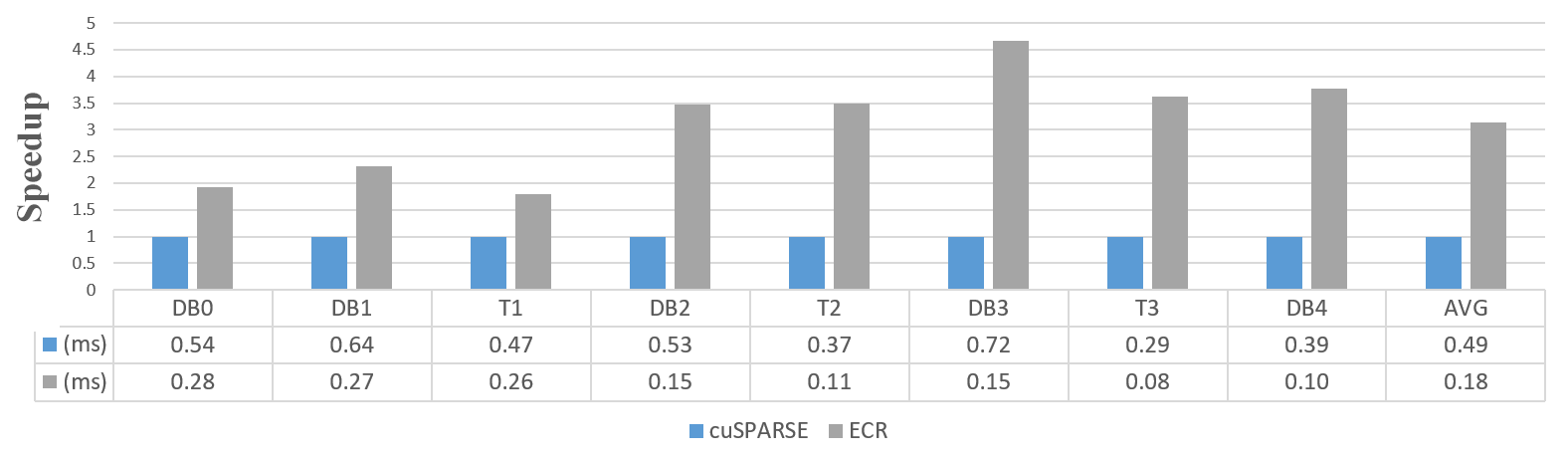}
			\label{cusparse_densenet}
		\end{minipage}
	}
	\subfigure[RegNetX-16GF]{
		\begin{minipage}[t]{0.48\linewidth}
			\centering
			\includegraphics[width=\textwidth]{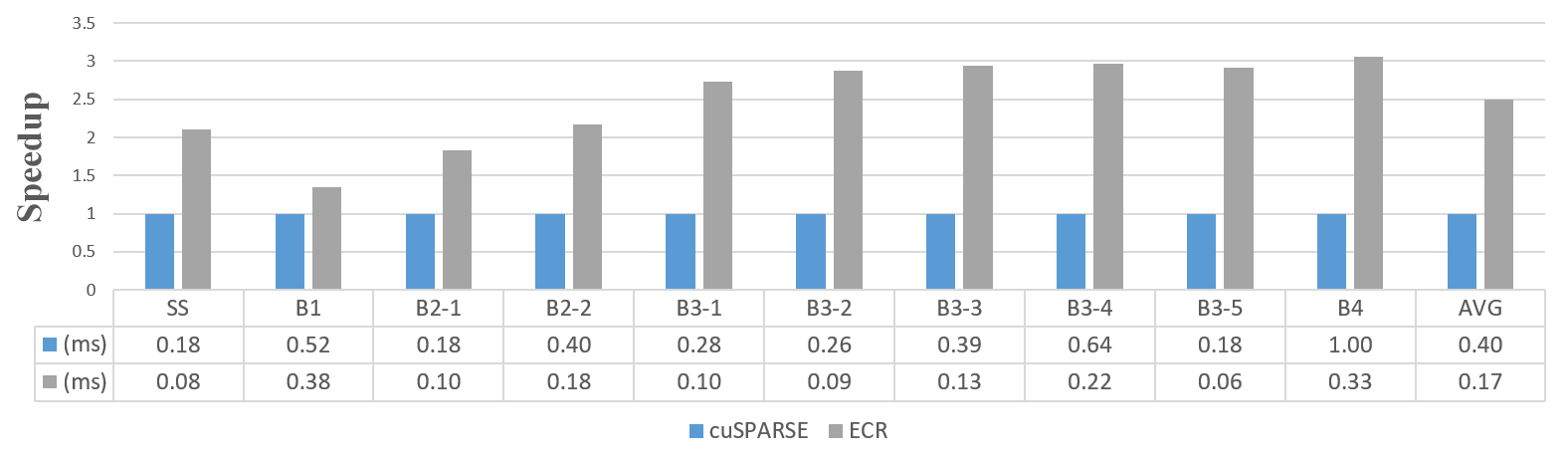}
			\label{cusparse_regnet}
		\end{minipage}
	}
\vspace{-10pt}
	\caption{Speed comparison between OCPA and cuSPARSE for VGG-19, ResNet-50, DenseNet-121, and RegNetX-16GF. The x-axis represents each layer of neural networks. The y-axis represents the acceleration effect of OCPA over cuSPARSE.}
	\label{cusparse}
\vspace{-20pt}
\end{figure}

We also use cuSPARSE to implement VGG-19, ResNet-50, DenseNet-121 and RegNetX-16GF. As cuSPARSE cannot compute pooling, OCPA only uses ECR as a comparison. 
Fig.\ref{cusparse_vgg} shows the speedup of OCPA over cuSPARSE for VGG-19. On average, the speedup of OCPA over cuSPARSE is $3.28\times$ for convolution layers in VGG-19. For the entire network of VGG-19, OCPA obtains 2.1× speedup.
As Fig.\ref{cusparse_resnet} shows, OCPA also gets better performance for ResNet-50 and achieves $3.3\times$ speedup over cuSPARSE for convolution layers on average. For entire ResNet-50, OCPA gets $1.83\times$ speedup.
As shown in Fig.\ref{cusparse_densenet}, the speedup of OCPA over cuSPARSE is 3.13× for convolution layers in DenseNet-121 on average. For the entire network of DenseNet-121, OCPA obtains 2.35× speedup. As Fig.\ref{cusparse_regnet} shows, OCPA also gets better performance for RegNetX-16GF and achieves 2.49× speedup over cuSPARSE for convolution layers on average. For the entire RegNetX-16GF, OCPA gets 1.35× speedup.
Compared with cuSPARSE, OCPA avoids redundant global memory accesses for extension and compression of feature maps, so OCPA can achieve better performance than cuSPARSE.
Note that we only use ECR of OCPA to compare with cuSPARSE, since cuSPARSE cannot compute pooling layers. Based on this situation, the speedup of OCPA over cuSPARSE is lower than cuDNN.

\subsection{Sensitivity analysis for sparsity}

\begin{figure}[!h]
	\subfigure[ECR]{
		\begin{minipage}{0.48\linewidth}
			\centering
			\includegraphics[width=\linewidth]{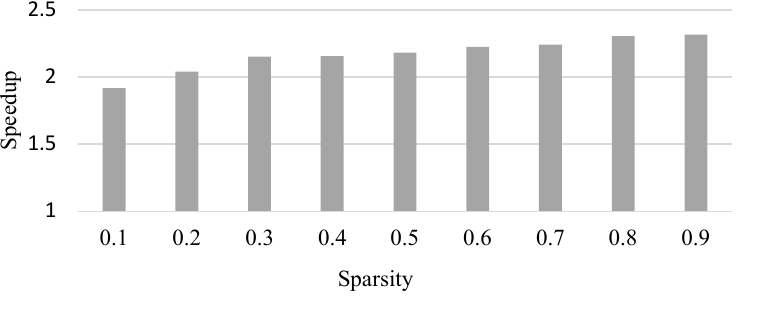}
			\label{ecr_sparsity}
		\end{minipage}
	}
	\subfigure[PECR]{
		\begin{minipage}{0.48\linewidth}
			\centering
			\includegraphics[width=\linewidth]{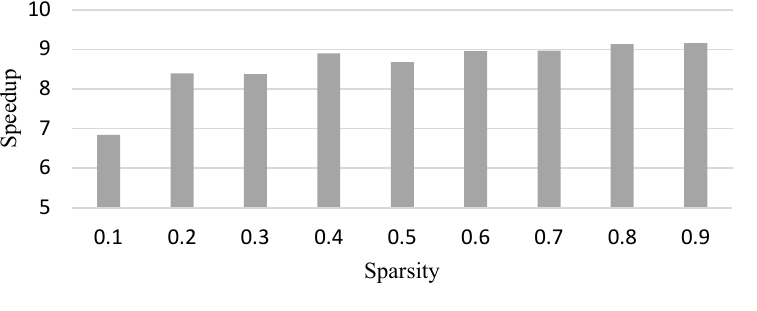}
			\label{pecr_sparsity}
		\end{minipage}
	}
 \vspace{-10pt}
	\caption{The speedup of ECR and PECR  over cuDNN (FAST) for feature maps with different sparsities.}
	\label{sparsity}
\vspace{-15pt}
\end{figure}

Both ECR and PECR use the sparsity of the feature map to accelerate CNN, so we try to analyze how the different sparsities can affect the speedup of ECR and PECR. Fig. \ref{ecr_sparsity} describes the speedup of ECR over cuDNN (FAST). We can see that the speedup  broadly becomes larger as the sparsity grows. 
Fig. \ref{pecr_sparsity} shows the speedup of PECR over cuDNN (FAST). We can see that the speedup also  broadly becomes larger as the sparsity increases. 
A larger sparsity leads to fewer computations for convolution so the performance of both ECR and PECR becomes better. On the other hand, the methods in cuDNN are not affected by the sparsity, so the speedup becomes larger.

\subsection{Sensitivity analysis for batch size}
\begin{figure}[!h]
	\subfigure[ECR]{
		\begin{minipage}[t]{0.48\linewidth}
			\centering
			\includegraphics[width=\linewidth]{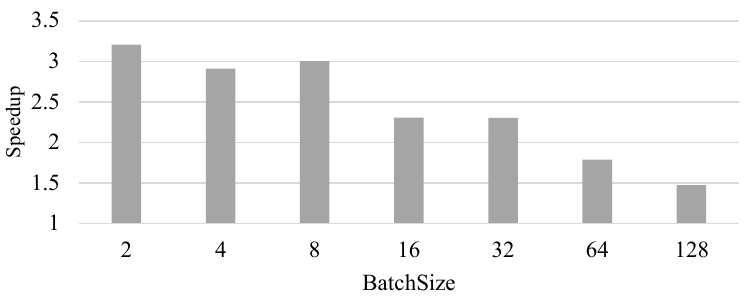}
			\label{ecr_batch}
		\vspace{-5pt}
		\end{minipage}
	}
	\subfigure[PECR]{
		\begin{minipage}[t]{0.48\linewidth}
			\centering
			\includegraphics[width=\linewidth]{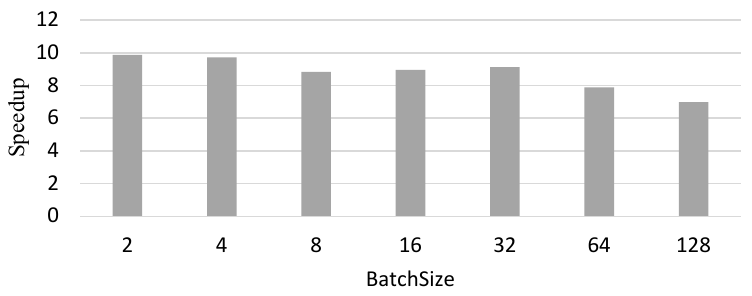}
			\label{pecr_batch}
		\vspace{-5pt}
		\end{minipage}
	}
 \vspace{-10pt}
	\caption{The speedup of ECR and PECR over cuDNN (FAST) for different batch sizes.}
	\label{batchsizes}

\vspace{-20pt}
\end{figure}

We also test how the different batch sizes can affect the speedup of ECR and PECR. Fig. \ref{ecr_batch} is the speedup of ECR over cuDNN (FAST). We can see that the speedup  broadly becomes smaller as the batch size becomes larger. However, we can still achieve about $1.5\times$ speedup even when the batch size is 128.
Fig. \ref{pecr_batch} is the speedup of PECR over cuDNN (FAST). We can see that the speedup also  broadly becomes smaller as the batch size becomes larger. However, we can still achieve more than $6\times$ speedup even when the batch size is 128. Both the time of our methods and the time of cuDNN increase as the batch size becomes larger. However, cuDNN may better exploit the parallelism of input data than our methods, so the speedup decreases.

\subsection{Sensitivity analysis for stride}
\begin{figure}[h]
	\centering
	\includegraphics[width=0.7\linewidth]{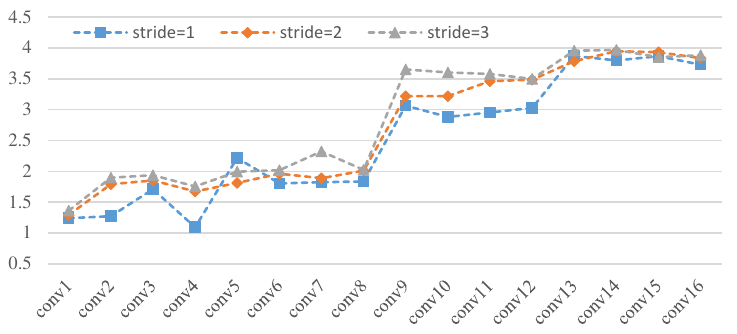}\\
 
	 \vspace{-5pt}
	\caption{Speedup of ECR over cuDNN (FAST) when convolution stride is 1, 2 and 3. 
	}
	\label{stride}
\vspace{-20pt}
\end{figure}

As shown in Fig. \ref{stride}, we also perform experiments with convolution strides of 1, 2, and 3 for all the convolution layers of VGG-19 using ECR. Our method can achieve the speedup of $2.51\times$ (stride is 1), $2.70\times$ (stride is 2) and $2.83\times$ (stride is 3) over cuDNN (FAST) on average. That means a larger speedup is obtained when the stride is larger. Both the time of our methods and the time of cuDNN decrease as the stride becomes larger. The performance of cuDNN is more easily affected by the increase of stride than our methods, so a larger speedup is obtained when the stride is larger. We also find that the speedup increases as the network goes deeper (from conv1 to conv16 in the figure). The main reason is that the sparsity becomes larger as the network goes deeper. 

\subsection{Overhead of data-format conversion}

\begin{figure}[!h]
	\subfigure[ECR]{
		\begin{minipage}[t]{0.4\linewidth}
			\centering
			\includegraphics[width=\linewidth]{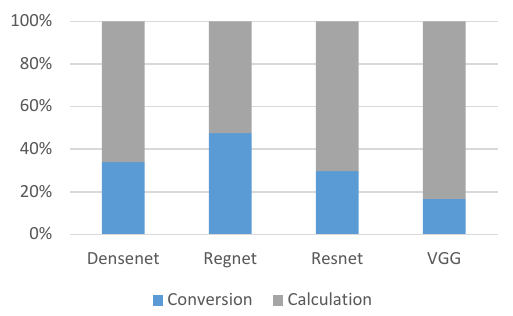}
			\label{ecr_conversion}
		\vspace{-5pt}
		\end{minipage}
	}
	\subfigure[PECR]{
		\begin{minipage}[t]{0.4\linewidth}
			\centering
			\includegraphics[width=\linewidth]{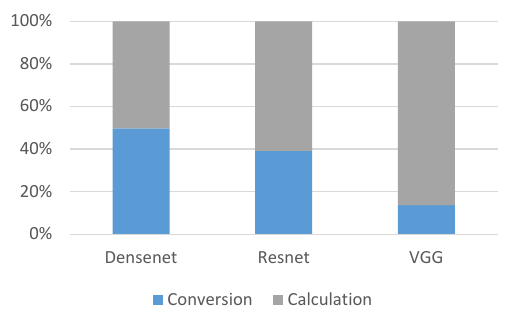}
			\label{pecr_conversion}
		\vspace{-5pt}
		\end{minipage}
	}
 \vspace{-10pt}
	\caption{Overhead of data-format conversion for ECR and PECR.}
	\label{data-format}

\vspace{-20pt}
\end{figure}

We carried out the experiment on data-format conversion overhead. As shown in Fig. \ref{ecr_conversion}, the time of data-format conversion takes up 34\% (DenseNet), 47\% (RegNet), 30\% (ResNet), and 17\% (VGG) of the convolution time on average for ECR. As shown in Fig. \ref{pecr_conversion}, the time of data-format conversion takes up 50\% (DenseNet), 39\% (ResNet), and 14\% (VGG) of the convolution and pooling time on average for PECR. Because RegNet only has a pooling layer at the end of the model, it is not implemented by PECR and not included in Fig. \ref{pecr_conversion}.
The proportion of conversion time for ECR and PECR has a lot to do with the size of the feature maps. A smaller feature map leads to a larger proportion.
The convolution layers of different models have different characteristics. For a model with a large number of convolution layers, the feature maps in the deeper convolution layer are smaller, so the proportion of format conversion time will be relatively large.

\section{Conclusion} \label{conclusion}

In this paper, two methods are proposed to optimize CNN on GPUs. First, the method based on ECR skips the computation for zero values in feature maps. 
Second, a PECR method is proposed not only to avoid computing zero values but also to compute convolution layer and pooling layer together, which effectively reduces the time both for transferring data between CPU and GPU and for loading data from global memory to shared memory on GPU.
Experimental results show that the proposed OCPA method can reach $1.97\times$,  $2.23\times$, $2.74\times$ and $1.58\times$ speedup over cuDNN (FAST) for VGG-19, ResNet-50, DenseNet-121 and RegNetX-16GF respectively. Over cuSPARSE, OCPA gets $2.10\times$, $1.83\times$, $2.35\times$ and $1.35\times$ speedup for the above four CNN models.

\vspace{-5pt}
\begin{acks}
This research is partially supported by NSF grants (CCF-2130688, CCF-1900904, and CNS-210705), Natural Science Foundation of Shandong Province (No. ZR2019LZH014 and ZR2022MF328), National Natural Science Foundation of China (No. 61602284 and No. 61602285), and the Funding for Study Abroad Program by the Government of Shandong Province.
\end{acks}

\bibliographystyle{ACM-Reference-Format}
\bibliography{acmart}










\end{document}